\documentclass{aa}

\usepackage{graphicx}
\usepackage{txfonts}
\usepackage[dvipsnames]{xcolor}
\usepackage{subcaption}
\usepackage{hyperref}
\hypersetup{
    colorlinks = True,
    citecolor= blue,
    linkcolor = blue,
    urlcolor = blue
}

\newcommand{\hii}{\ion{H}{ii}\xspace}
\newcommand{\quotes}[1]{`#1'}

\newcommand{\siit}{[\ion{S}{ii}]$\lambda6716,6731$} 
\newcommand{\hb}{H$\beta$} 
\newcommand{\oiii}{[\ion{O}{iii}]} 
\newcommand{\oii}{[\ion{O}{ii}]} 
\newcommand{\oi}{[\ion{O}{i}]} 
\newcommand{\ha}{H$\alpha$} 
\newcommand{\nii}{[\ion{N}{ii}]} 
\newcommand{\sii}{[\ion{S}{ii}]} 
\newcommand{\siii}{[\ion{S}{iii}]} 
\newcommand{\ariii}{[\ion{Ar}{iii}]}
\newcommand{\hei}{\ion{He}{i}\xspace}

\begin{document}

\title{Classifying spectra of emission-line regions with neural networks}

\subtitle{An application to integral field spectroscopic data of M33}

\author{Caterina Bracci\inst{1, 2}
        \and
        Francesco Belfiore\inst{2}
        \and
        Michele Ginolfi\inst{1,2}
        \and
        Anna Feltre\inst{2}
        \and
        Filippo Mannucci\inst{2}
        \and
        Alessandro Marconi\inst{1,2}
        \and
        Giovanni Cresci\inst{2}
        \and
        Elena Bertola\inst{2}
        \and
        Alessandro Bombini\inst{3,4}
        \and
         Matteo Ceci \inst{1,2}
        \and
        Cosimo Marconcini \inst{1,2}
        \and
        Bianca Moreschini \inst{1,2}
        \and
        Martina Scialpi \inst{1,2,5}
        \and
        Giulia Tozzi \inst{6}
        \and
        Lorenzo Ulivi\inst{1,2,5}
        \and
        Giacomo Venturi \inst{7,2}
    }
          
    \institute{
    Dipartimento di Fisica e Astronomia, Università di Firenze, Via G. Sansone 1, I-50019, Sesto F.no (Firenze), Italy
    \and
    INAF - Osservatorio Astrofisico di Arcetri, Largo E. Fermi 5, I-50125, Florence, Italy\label{arcetri}
    \and 
    INFN - Istituto Nazionale di Fisica Nucleare (INFN), Via Bruno Rossi 1, 50019 Sesto Fiorentino (FI), Italy 
    \and
    ICSC - Centro Nazionale di Ricerca in High Performance Computing, Big Data \& Quantum Computing, Via Magnanelli 2, 40033 Casalecchio di Reno (BO), Italy
    \and
    University of Trento, Via Sommarive 14, I-38123 Trento, Italy 
    \and
    Max-Planck-Institut für Extraterrestrische Physik (MPE), Gießenbachstraße 1, 85748 Garching, Germany 
    \and
    Scuola Normale Superiore, Piazza dei Cavalieri 7, I-56126 Pisa, Italy\\ 
    }


\abstract{
Emission-line regions are key to understanding the properties and evolution of galaxies, as they trace the exchange of matter and energy between stars and the interstellar medium (ISM). In nearby galaxies, individual nebulae can be resolved from the background, and classified through their optical spectral properties. Traditionally, the classification of these emission-line regions in \hii regions, planetary nebulae (PNe), supernova remnants (SNRs), and diffuse ionised gas (DIG) relies on criteria based on single or multiple emission-line ratios. However, these methods face limitations due to the rigidity of the classification boundaries, the narrow scope of information they are based upon, and the inability to take line-of-sight superpositions of emission-line regions into account.
In this work, we explore the use of artificial neural networks to classify emission-line regions using their optical spectra. Our training set consists of simulated optical spectra, obtained from photoionisation and shock models, and processed to match observations obtained with the MUSE integral field spectrograph at the ESO/VLT. We evaluate the performance of the network on simulated spectra exploring a range of signal-to-noise (S/N) levels, different values for dust extinction, and the superposition of different nebulae along the line of sight. At infinite S/N the network achieves perfect predictive performance, while, as the S/N decreases, the classification accuracy declines, reaching an average of $\sim80\%$ at S/N(\ha)=20.
We then apply our model to real spectra from MUSE observations of the Local Group galaxy M33. 
The network provides a robust classification of individual spaxels, even at low S/N, 
identifying \hii regions and PNe and distinguishing them from SNRs and diffuse ionised gas, while identifying overlapping nebulae. Moreover, we compare the network's classification with traditional diagnostics, finding a satisfactory level of agreement between the two approaches.
We identify the emission lines that are most relevant for our classification tasks using activation maximisation maps. 
In particular, we find that at high S/N the model mainly relies on weak lines (e.g. auroral lines of metal ions and He recombination lines), while at the S/N level typical of our dataset the model effectively emulates traditional diagnostic methods by leveraging strong nebular lines.
We discuss potential future developments focused on deriving segmentation maps for \hii\ regions and other nebulae, and the extension of our method to new datasets and instruments.
}

\keywords{ Methods: machine learning -- Methods: data analysis -- Galaxies: ISM --
   ISM: general --
   ISM: \hii regions}
\titlerunning{Classifying spectra of emission-line regions with neural networks}
\authorrunning{C. Bracci}
\maketitle
%

\begin{figure*}[t]
    \centering
    \includegraphics[width=1\linewidth]{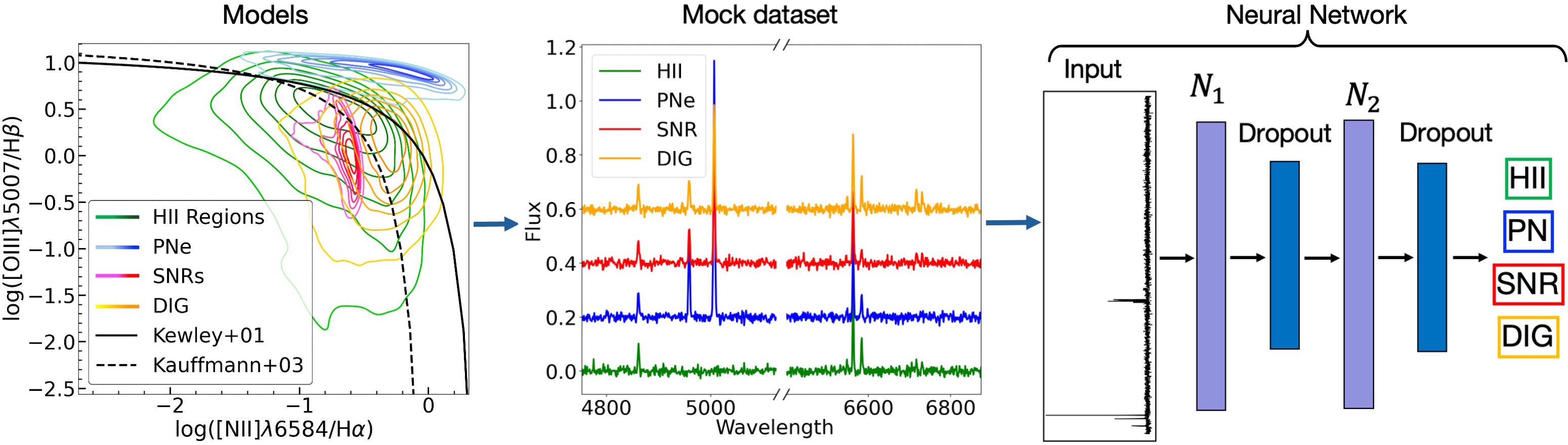}
    \caption{Schematic representation of the workflow presented in this paper. The diagram illustrates the key steps and processes involved in the analysis, from choosing the emission-line region models to the generation of the simulated dataset of spectra and the neural network's classification. The leftmost image illustrates the model selection phase, depicting the distribution of our \hii region, PN, SNR, and DIG models on a BPT diagnostic diagram \citep{Baldwin1981}. The central plot provides a representative simulated spectrum for each of the four classes. The rightmost panel presents a schematic of the neural network architecture, where the simulated spectra serve as input, and the network is trained to classify each spectrum into its corresponding nebular type.}
    \label{fig:scheme}
\end{figure*}

\section{Introduction}
\label{intro}

Ionised nebulae act as signposts of diverse astrophysical processes, including those governing the life cycle of stars, and feedback of stars into the interstellar medium (ISM).
\hii regions, ionised by young, massive O and B stars, for example, provide key information on chemical composition, star formation and feedback in galaxies through their emission-line-rich optical spectra \citep{Stasinska2004, Krumholz2009,Kreckel2018,Barnes2020,Santoro2022,Sanders2024}.
Planetary nebulae (PNe), on the other hand, are formed from gas ejected by intermediate-mass stars which ionise the surrounding material as they evolve from the asymptotic giant branch into white dwarfs. PNe have been used as tracers of galaxy dynamics, chemical composition, and to measure extragalactic distances via the planetary nebula luminosity function \citep{Acker1992, Maciel2013, Roth2021, Scheuermann2022}.
Supernova remnants (SNRs), are instead ionised by collisional shocks rather than photoionisation \citep{Fesen1985, Iben1984, Maciel2013}. Supernovae are amongst the main sources of feedback and chemical enrichment into the ISM. Therefore, understanding the evolution of SNRs and estimating the rate of supernovae in galaxies is essential for the study of the energetics and chemical evolution of the ISM \citep{Blair2004, Kim2018, Koo2020}.

In nearby galaxies, it is possible to spatially resolve \hii regions and other ionised nebulae, therefore morphologically isolating them from the background emission of the diffuse ionised gas (DIG). 
Traditional methods for identifying and classifying ionised regions rely on narrow-band imaging and single or multiple emission-line ratios.
At solar metallicity, \hii regions are generally traced by the Balmer \ha\ line, which is the most prominent in their emission-line-rich spectra. The most commonly used diagnostic diagrams, such as the Baldwin-Phillips-Terlevich (BPT, \citealt{Baldwin1981, Veilleux1987}) diagram, are based on ratios of pairs of strong optical emission lines (e.g. \nii$\lambda$6584/\ha, \oiii$\lambda$5007/\hb) to separate \hii\ regions from other types of nebulae \citep{Kreckel2019, DellaBruna2020,Santoro2022}.

Overall, traditional methods for classifying ionised nebulae 
(see, for example, \citealt{Maiolino2019})
come with significant limitations. 
Most of them are designed to identify a single type of nebula at the time, rather than enabling comparative classification.
Moreover, these criteria also rely on sharp empirical or theoretical boundaries \citep{Kewley2001, Kauffmann2003}, which can lead to misclassification of objects with low S/N ratios or those near the edge of these limits. 
Overlap between different nebula types in the line-ratio space further complicates the classification \citep{Frew2010, Kopsacheili2020}. 
These difficulties mean that such a classification often requires visual inspection to confirm morphological details or to link nebulae to their ionising sources \citep{Rousseau-Nepton2018, Kreckel2019}, which is particularly impractical in the context of large datasets. 

\quotes{Cloud-scale} integral field spectroscopic (IFS) surveys, with spatial resolutions better than $100$ pc, enable us to map and resolve individual structures involved in the star formation process (with sizes from a few parsecs to $\sim100$ parsecs; \citealt{Sanders1985,Oey2003}). These advances have revolutionised ISM studies, enabling detailed exploration across diverse galaxy samples 
(e.g. SIGNALS, \citealt{Rousseau-Nepton2019}; MAGNUM, \citealt{Mingozzi2019}; PHANGS, \citealt{Emsellem2022}), while emphasising the need for efficient classification methods to manage the increasingly large and complex datasets they provide.

For these reasons, in this work, we explore the use of machine learning (ML), and neural networks in particular, for the classification of emission-line regions. Several studies have applied ML techniques to identify different regions in galaxies (e.g. \citealt{Rhea2023,Baron2024,Belfiore2024}) and to determine their astrophysical properties (e.g. \citealt{Ucci2017,Kang2023, Ginolfi2025}).
Recently, \cite{Rhea2023} classified \hii\ regions, PNe, SNRs, and DIG by training a fully connected neural network on three key line ratios present in the optical spectra of these regions: \oiii$\lambda5007$/H$\beta$, \nii$\lambda6583$/H$\alpha$, and \sii/\ha. 
The line ratios for the training were obtained from photoionisation and shock models for each class of nebulae.

In Fig. \ref{fig:scheme}, we show a schematic view of the approach introduced in this paper, which expands the approach of \cite{Rhea2023}. Starting from photoionisation and shock models, we created a dataset of simulated optical spectra. 
We then fed the spectra (as opposed to a table of individual line ratios) to the network and trained a model to classify the different nebular spectra.
This approach is advantageous because it allows us to self-consistently apply our classification algorithm to regions where certain emission lines are undetected. 
It also enables the network to exploit the additional information provided by weaker emission lines in regions of higher S/N, going beyond the more common optical diagnostics.  

We also account for the potential superposition of nebulae along the line of sight by training our model on simulated spectra, which are constructed as linear combinations of different nebular classes. This approach not only addresses the complexity introduced by overlapping regions but also helps to reproduce the spectra of individual emitting regions better, which are typically more complex than those described by a single-cloud photoionisation model. 
Overall, our approach aims to enable the classification of nebulae with contamination from other nebulae (such as the superposition along the line of sight) and low S/N.

The Local Group galaxies are the ideal laboratory to develop, validate, and test classification methods for emission-line regions because these galaxies are close enough to allow us to observe and spatially resolve individual nebulae. 
M33 is the third largest galaxy in the Local Group and the second closest spiral to the Milky Way, located at a distance of approximately 840 kpc \citep{Gieren2013, Breuval2023}. Its proximity, combined with the data quality provided by MUSE, makes it an excellent target for testing the performance of our model on real data.

In this work, therefore, we present in Sec. \ref{sec:data} the M33 MUSE data. In Sec. \ref{sec:methods}, we describe the generation of the simulated spectra and our neural network model architecture. In Sec. \ref{sec:test}, we evaluate the performance of the model on the simulated spectra, while in Sec. \ref{sec:application} we discuss the classifications of the MUSE M33 dataset. Finally, in Sec. \ref{sec:discussion}, we discuss our results using interpretability techniques and compare them with traditional methods, while in Sec. \ref{sec:conclusions}, we present a summary of our results.


\section{Data}
\label{sec:data}

In this work, we use MUSE IFS data of the nearby galaxy M33 and build simulated spectra to match the observational characteristics of this dataset. We present in the following sections the data acquisition and the processing methods used in this study.

The observations of M33 used in this work were obtained using MUSE in its seeing-limited, wide-field mode, and are part of the ESO programme ID 109.22XS.001 (PI: G. Cresci). 
A mosaic of 24 $1'\times1'$ MUSE pointings covered a mostly contiguous region of $3'\times8'$ (0.7 $\times$ 1.9 kpc$^2$) along the southern major axis of the galaxy, with an overlap of $2''$ between adjacent pointings. The average seeing during observations was $\sim 0.7''$, corresponding to $\sim 2.9$ pc at the distance of M33 of $840$ kpc.
The MUSE data have a spectral resolution ($R = \lambda/\Delta\lambda$) ranging from 1770 at 4800~$\AA$ to 3590 at 9300~$\AA$.

The observation and data reduction strategy is discussed in detail in Feltre et al. (in prep.), and was performed with the {\tt pymusepipe} software framework\footnote{\url{https://pypi.org/project/pymusepipe/}}, a customised python wrapper to the MUSE data reduction pipeline \citep{Weilbacher2020} developed by the PHANGS team \citep{Emsellem2022} and optimised for processing large mosaics of extended objects.  
The final datacube, created by aligning the 24 pointings, contains $\sim 2.3 \times 10^6$ spectra, each consisting of 3682 wavelength channels.
Feltre et al. (in prep.) also derived emission-line flux maps from the MUSE M33 datacube. These maps allowed us to compare the predictions from the ML models to standard diagnostics based on line ratios of optical lines. 

\section{Methods}
\label{sec:methods}

\begin{figure*}
    \centering
    \includegraphics[width=0.99\linewidth]{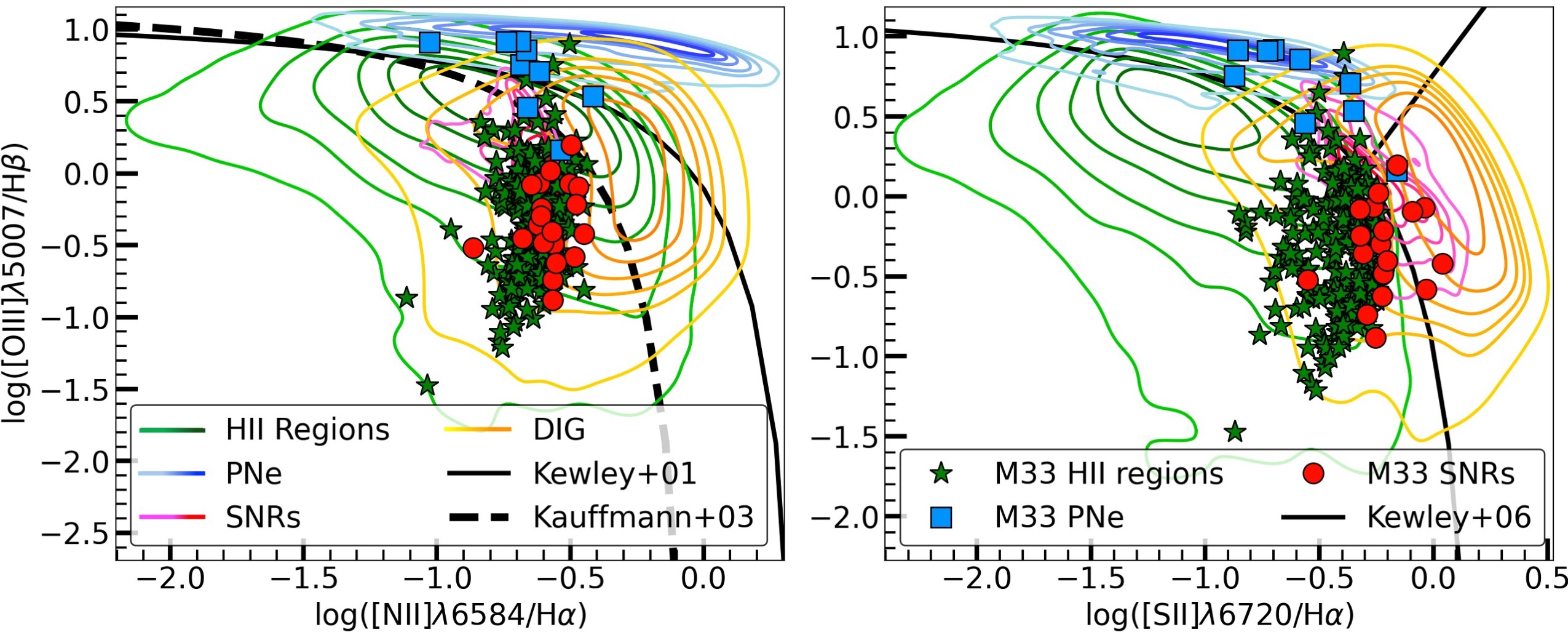}
    \caption{Two diagnostic diagrams (\nii-BPT, \citealt{Baldwin1981}; \sii-BPT, \citealt{Veilleux1987}) illustrating the relative positions of the emission-line region models and the observed nebulae in M33 across three different line ratio spaces. Green stars mark the identified M33 \hii regions, blue squares the M33 PNe, and red circles the M33 SNRs. Similarly, green contours show the \hii regions models, blue ones the PN models, pink-red ones the SNR models, and yellow-orange ones the DIG models. On the left, the \nii-BPT diagram \citep{Baldwin1981} includes the \cite{Kewley2001} and \cite{Kauffmann2003} (dashed line) diagnostic lines. On the right, the \sii-BPT diagram includes the \cite{Kewley2006} diagnostic line.}
    \label{fig:bptplots}
\end{figure*}

\subsection{Photoionisation and shock models}
\label{sec:mmm}

We created a set of synthetic emission-line region optical spectra, exploiting predictions from photoionisation and shock models for the types of nebula of interest. In particular, we selected grids of simulations from the Mexican Million Models database (3MdB, \citealt{Morisset2015}, \citealt{Alarie2019}), computed using the \texttt{Cloudy v13} photoionisation code \citep{Ferland2013} or the shock modelling code \texttt{MAPPINGS V} \citep{Sutherland2018}.

We took inspiration from \cite{Congiu2023} and \cite{Belfiore2024} in selecting a suitable set of \hii regions, PNe, and SNRs models and parameters from those available in 3MdB. A detailed description of the model selection is provided in Appendix \ref{sec:AppA}. We added DIG models and refined the parameter selection for the models of the other three nebulae to better match those of observed nebulae in M33 (see Appendix \ref{sec:regions}) in the \nii\ and \sii-BPT diagrams. 

Fig. \ref{fig:bptplots} shows the distribution of the final set of models in the \nii-BPT and \sii-BPT diagrams. These distributions are compared with the integrated spectra from the M33 nebulae, showing a reasonable overlap.
The diagnostic lines from \cite{Kewley2001}, \cite{Kauffmann2003}, and \cite{Kewley2006} are included to distinguish between different ionisation mechanisms, such as star formation and shocks, and roughly match the expected location with respect to the different model classes.

\subsection{Generation of mock spectra}
\label{sec:mockspec}

We used the photoionisation and shock models presented in Sect. \ref{sec:mmm} to generate \quotes{realistic} \hii regions, PNe, SNRs, and DIG mock spectra, matching the wavelength range, sampling, spectral resolution, noise, and other physical parameters of the MUSE data for M33. In this Section we describe the process in more detail.

We first restricted our analysis to emission lines that are covered by the wavelength range of MUSE and, for practical reasons, considered only a set of emission lines which can typically be detected in our dataset and in spectra of typical \hii\ regions (e.g. \citealt{Berg2020}). 
Line intensities were then normalised to the \ha\ intensity.

We modelled each emission line with a Gaussian function centred at its rest-frame wavelength shifted to the systemic velocity of M33. 
The velocity dispersion for each emission line was computed as the quadrature sum of the MUSE line spread function and the intrinsic velocity dispersion of the ionised nebula. We adopted the MUSE line spread function from \cite{Bacon2017}, which varies from $\sigma \sim 80 ~\rm km~s^{-1}$ at the blue end of the spectrum to $\sim 40  ~\rm km~s^{-1}$ at the red end. The intrinsic velocity dispersion, on the other hand, was randomly selected for each spectrum in the range $[30-50] ~\mathrm{km~s^{-1}}$, which represents the expected range for PNe and \hii\ regions \citep{Magrini2016, Kewley2019}. 
SNRs are expected to have larger velocity dispersion (which can exceed $\sim100 ~\mathrm{km~s^{-1}}$ \citealt{Li2024}), especially when unresolved. However, such behaviour is not evident in our M33 data. We therefore did not use a different distribution of velocity dispersion for SNRs and DIG.

We added random Gaussian noise to the spectra to mimic the effect of observational noise. For the purpose of testing our classification models, we consider two different noise levels, $\mathrm{S/N} =20$ and $\mathrm{S/N}=100$. Since our mock spectra are normalised to an \ha\ flux of one, these values correspond to S/N values on the \ha\ line flux. These values are chosen because an \ha\ S/N = 20 corresponds to the 20$\rm ^{th}$ percentile of the S/N distribution in the M33 datacube, while $\mathrm{S/N}=100$ corresponds to the 99$\rm ^{th}$ percentile. From this point on, we refer to the $\mathrm{S/N} =20$ and $\mathrm{S/N}=100$ model set as \quotes{low S/N} and \quotes{high S/N} respectively. Additionally, we refer to the dataset with no added noise as \quotes{infinite S/N}.

We also model dust attenuation assuming the \cite{Cardelli1989} Galactic extinction law as revised by \cite{ODonnell1994}, with $R_V = 3.1$. We randomly extract $E(B-V)$ values for each spectrum from a uniform distribution, ranging from $E(B-V)=0$ to $0.4$, where the latter value corresponds to the 90$\mathrm{^{th}}$ percentile of the distribution of the $E(B-V)$ values in the M33 MUSE datacube (see Sec. \ref{sec:ifsclassif}). 
Similarly to the addition of noise, this process adds variety to the dataset. For testing, we consider a dataset with dust attenuation and S/N=100, which we refer to as \quotes{high S/N $+$ extinction}.

\subsection{Preprocessing of the mock spectra}

Before feeding the mock spectra into our ML model, we experimented with different normalisation methods. Standardisation is a common preprocessing step in ML, where features are rescaled to have a mean of zero and a standard deviation of one, ensuring that the model does not favour features with larger numerical ranges. 
We also tested min-max normalisation, in which we scale data to a fixed range, such as [0, 1].
Both these methods are typically performed across all instances in the dataset for each feature (e.g. each wavelength channel in our spectra) to ensure consistent scaling of input features. 
In our case, we also explored normalising each input spectrum independently along the wavelength dimension. This approach ensures that the relative differences within each spectrum are preserved while removing variations in scale or offset between spectra.  

We found that the performance of the network did not depend on the choice of normalisation so we opted for the standardisation of individual spectra across wavelength, which has the advantage of being trivial to apply to observational data and ensures that the spectra remain interpretable. 
Afterwards, we divided our mock spectra into training (70\%), validation (15\%) and testing (15\%) subsets. 

\subsection{Neural network architecture}

We tested different model architectures including both dense fully connected neural networks and 1D convolutional neural networks (CNN). While in a fully connected network every neuron in each layer is connected to every neuron in the next layer, a CNN uses convolutional layers to better detect features, reducing the number of parameters compared to dense connections. The key difference is therefore that the fully connected networks treat all input features (in our case wavelength channels) equally, while the convolutional ones exploit local relationships in the data. We found that the performances of our both models were similar, so we focus here on the results obtained using the fully connected network.

Our default neural network model consists of a fully connected network with two dense hidden layers interspersed with dropout layers. 
For the training with the infinite S/N dataset, we use a first hidden dense layer with 16 units, followed by a dropout layer (with a dropout rate of 5\%), a second dense layer with four neurons, again followed by a 5\% dropout layer, followed by the output layer. The output layer consists of four neurons with a softmax activation to calculate the probabilities of each output class. For the task of classifying noisy spectra we use a first dense layer of 128 neurons, and a second layer of 64. Each is followed by a dropout layer (with 10\% dropout for the high S/N cases and 20\% dropout for the low S/N case). These are followed by the usual output layer. A visual representation of the network's architecture can be viewed in Fig. \ref{fig:scheme}, right panel.

The models are fit with a batch size (the number of instances in the mini-batches used in the training) of 256. The learning is stopped with the \texttt{EarlyStopping} callback, after five epochs of patience in which the loss measured on the validation set does not decrease. We trained the model with cross-entropy loss and the Adam (Adaptive Moment Estimation) optimizer algorithm \citep{Kingma2014}.
We build the networks using \texttt{keras} (\texttt{v2.14.0}; \citealt{chollet2015keras}) from \texttt{tensorflow} (\texttt{v2.14.0}; \citealt{Abadi2016}), implemented in \texttt{python} (\texttt{v3.9.7}).

\section{Testing on mock spectra}
\label{sec:test}

In this section, we discuss the performance of the model on a subset of mock spectra by comparing its classifications with the known input labels. In particular, we evaluated the model using confusion matrices, which compare the input and predicted classes, and using the f1-score, which is the harmonic mean of precision and recall. 
In multi-class classification, precision for each class measures the proportion of true positives among all predicted positives for that class, while recall for each class measures the proportion of true positives among all positives for that class. The overall f1-score was computed as the unweighted average of the f1-scores for all classes since the number of instances is the same for each class.

\subsection{The effect of noise and extinction}
\label{sec:noiseext}

\begin{figure}[t]
    \centering
    \includegraphics[width=\linewidth]{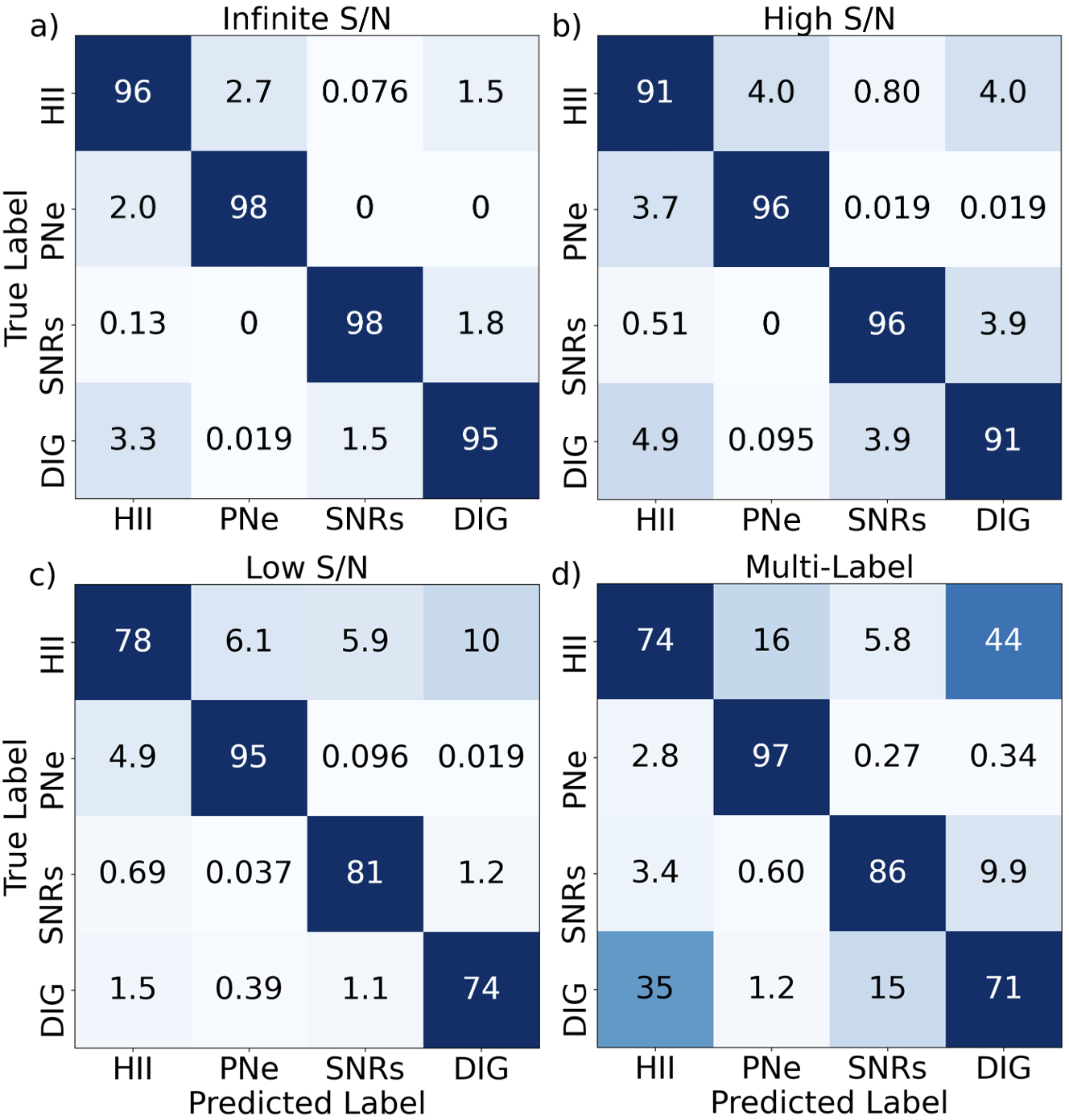}
    \caption{Confusion matrices comparing the networks' predicted classification with the true classification for four different test mock datasets. The values shown are percentages. Top row: confusion matrices for the network trained with the single-label high S/N dataset without extinction (panel a) and for the network trained with the high S/N dataset with extinction (panel b). Bottom row: confusion matrices for the network trained with the low S/N dataset without extinction (panel c) and for the network trained with the multi-label dataset (panel d).}
    \label{fig:cm_all}
\end{figure}

We compare performance metrics for models trained with the four different mock datasets described in Section \ref{sec:mockspec}: infinite S/N, high S/N, high S/N $+$ extinction, and low S/N. 
For the infinite S/N case, the confusion matrix is perfectly diagonal (not shown), as the model is able to recognise the correct label for each instance of the test set. 
As visible in Fig. \ref{fig:cm_all} and quantified in Table \ref{tab:performance}, the model's performance deteriorates with the addition of noise and extinction effects, with f1-scores decreasing to 0.82 for the low S/N case. 
However, in the high S/N and high S/N + extinction cases, where the datasets consist of high-quality spectra, f1-scores above $90\%$ can be achieved.

The PN class is consistently more accurately recovered, whereas the classification performance for the \hii region and DIG classes declines more significantly when adding noise and extinction (to $78\%$ and $74\%$ respectively in the low S/N case).
Overall, we also find a higher misclassification rate between \hii regions and DIG spectra, \hii regions and PNe spectra, and SNR and DIG spectra.

\begin{table}[tb]
    \centering
    \caption{f1-scores on the test set of the ANN model trained on different mock datasets.}
    \begin{tabular}{cccc}
    \hline\hline
    Mock dataset & S/N & E($B-V$) & Test f1-score \\
    \hline
    Infinite S/N & $\infty$  & 0 & 1.00 \\
    High S/N & 100 & 0 & 0.97 \\
    High S/N  $+$ extinction & 100 & $[0-0.4]$ & 0.94 \\
    Low S/N & 20 & 0 & 0.82 \\
    Multi-label & 100  & $[0-0.4]$  &  0.82 \\
    Single spaxel & $<37>$* & $<0.16>$* & 0.80 \\
    Integrated nebulae & 370 & $<0.15>$* & 0.89 \\
    \end{tabular}
   \tablefoot{The single spaxel and integrated nebulae entries are related to the models trained for the application to the M33 data. The f1-scores measured for both these two models and the multi-label one consider only the instances with a contribution from the prevalent class >$50\%$. The \quotes{*} marks average values of the S/N or extinction distributions that have been tailored to match that of the real data, as described in Sec. \ref{sec:tailoring}.}
    \label{tab:performance}
\end{table}

\subsection{Mock spectra of nebular line-of-sight superpositions (`multi-label')}
\label{sec:mlspec}

Realistic observations along a line-of-sight might contain a superposition of nebulae of different classes and/or be contaminated by foreground and background diffuse emission. In this section we explore the ability of our ML model to recover the contributions of different components. To do this, we built a new set of mock spectra as a weighted sum of spectra from individual classes. We refer to the models obtained by linear combination as `multi-label' and the original single-class models for \hii\ regions, PNe, SNR, and DIG as `single-label'.

We built multi-label models by generating weighted sums of the line intensities of four single-label models, one for each of the four different classes. 
For example, a multi-cloud model described by the label [0.2, 0.6, 0.1, 0.1] is constituted by a random \hii region model scaled by 0.2, a PN model scaled by 0.6, a SNR model scaled by 0.1, and a DIG model also scaled by 0.1. 
Because of their normalisation, these weights can be interpreted as probabilities. Therefore, we also set up the problem of the recovery of these weights as a multi-label classification problem.

In total, we generate $200\,000$ \quotes{multi-label} models, selecting the single-label models randomly but making sure that each model is taken into account a few times. The number of multi-label instances significantly exceeds the number of \quotes{single-label} ones in this dataset (they are 1/6 of the total number of models), because the task of classifying the multi-label spectra is more complex than that of classifying single-label spectra as the multi-label linear combinations can be extremely diverse among themselves. Therefore the network will need many examples to learn to correctly classify these instances.
Starting from these emission-line region models, the mock spectra are created as described in Sec. \ref{sec:mockspec} with a S/N=100 and $E(B-V)=0-0.4$ (matching the characteristics of the low-noise dataset with extinction).
We chose the same network architecture as the one used for the classification of noisy spectra (Sec. \ref{sec:methods}). 

To evaluate the network performance, we consider a test set of spectra where the prevalent contribution is higher than $50\%$, to mimic cases where a nebula, dominating the emission, is subject to some degree of contamination. The \quotes{true label} is taken as the class with the highest contribution in each spectrum's label and the \quotes{predicted label} is taken as the class with the highest contribution (i.e. with the highest probability) as predicted by the network. The results are shown in Fig. \ref{fig:cm_all}, panel d. As for the single-label cases discussed above, we find that the PNe class is predicted most accurately, while the \hii regions and DIG classes have lower accuracy. In particular, the network struggles to recover the true label of regions where the \hii region class is the one with the highest contribution. Finally, we measure an f1-score of 0.82. This value is comparable with that of the low S/N single-label dataset (Sec. \ref{sec:noiseext}), even if the training sample has a higher S/N of 100. However, this is justifiable considering that the task has been considerably complicated by the mixture of different nebular models.

For a more comprehensive view of the network's performance, we measure the difference, for each class, between the input weight and the model probability prediction of each spectrum of the test set. Then we calculate the median of the absolute values of the difference for each class. 
The median errors on the weights of the four classes are 8.9, 5.3, 7.2, and 8.2\% respectively for \hii regions, PNe, SNRs, and DIG. Since the sum of the weights is normalised to one, these numbers can be interpreted as median fractional errors. The network therefore generally achieves good performance, with fractional errors on class contributions of less than 9\%. The PNe contribution is the better predicted one, while the contributions from \hii regions and DIG are predicted a little less accurately.

\begin{figure}
    \centering
    \includegraphics[width=\linewidth]{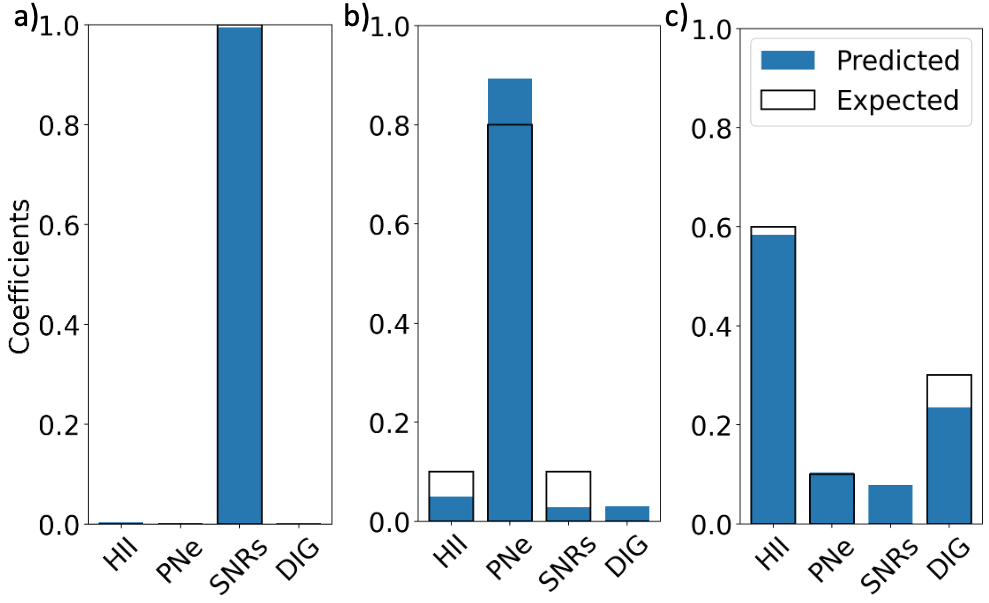}
    \caption{Bar plots showing the median error made by the network trained on the multi-label dataset on each class contribution for three example instances. The coloured bars represent the network's prediction, while the empty bars are the true labels. In panel a, we show the classification of a single-label PNe spectrum, whose coefficients are accurately predicted. In panel b, we show the results for a multi-label instance, with a prevalent contribution of DIG $>60\%$. In panel c, we show the results for another multi-label instance, with a prevalent contribution of \hii region $=60\%$.}
    \label{fig:mlbarpltexamples}
\end{figure}

Fig. \ref{fig:mlbarpltexamples} shows the discrepancy between the input weight and the model prediction for four example instances of the test set with different combinations of the four nebulae. 
The plot in panel a) shows a single-label SNR spectrum, whose coefficients are well predicted by the network.
The b) and c) panels show two multi-label cases in which the prevalent class has a contribution higher than 0.5: we notice that the network retrieves the prevalent class, even if with some error on the estimate ($<0.1$).

\section{Application to observational data of M33}
\label{sec:application}

\subsection{Tailoring of the models to observed data}
\label{sec:tailoring}

To apply our neural network to real observational data, we train new models using mock spectra tailored to match the specific S/N and extinction distributions of our M33 dataset.

Such an adjustment is necessary because variations in S/N significantly influence the model's ability to classify or predict accurately (e.g. \citealt{Kang2023, Ginolfi2025}).
In particular, if the S/N and extinction properties of the training data differ significantly from those of the test data, the model’s performance can degrade due to a mismatch in learned features. For instance, a model trained on high S/N spectra may struggle to classify low S/N data accurately and vice versa, as the noise characteristics can dominate or obscure the spectral features critical for classification. 

For example, when low S/N instances (S/N=20) are classified using our model trained on high S/N spectra (S/N=100), a significant drop in performance is observed, with an f1-score of 0.56 compared to 0.82 (as shown in Table \ref{tab:performance}), achieved when the classification is performed using a model trained on data with the same S/N.
Therefore, in Sec. \ref{sec:tailoring_spaxels} and \ref{sec:tailoring_integrated}, we present the process of tailoring the models to classify the spectra in each spaxel of the M33 field. In Appendix \ref{sec:regions}, we explore a different model tailored to match the characteristics of integrated nebulae rather than single spaxels.

\subsection{The `single spaxel' model}
\label{sec:tailoring_spaxels}

To apply our ML model to the MUSE M33 cube, we first pre-processed the input datacube, enhancing the S/N of the spectra by applying a spatial Gaussian smoothing with a kernel of $\sigma = 1$ spaxel.
We then generated a new multi-label mock spectra dataset with a S/N distribution matching that of the data. In particular, we fit the \ha\ S/N distribution from the M33 datacube with a lognormal distribution (mean $\sim37$, $\sigma\sim22$). We then sampled suitable noise values from this distribution to generate mock spectra.

We recreated in our mock spectra the distribution of $E(B-V)$ inferred from the Balmer decrement (\ha/\hb) from each spectrum of the MUSE data. In particular, we calculated the dust attenuation applying a S/N > 5 cut on \ha\ and \hb\ fluxes, assuming the case B theoretical ratio of \ha/\hb = 2.87 and adopting the \cite{Cardelli1989} Galactic extinction law as revised by \cite{ODonnell1994}, with $R_V = 3.1$. 
We find the distribution of $E(B-V)$ for the individual spaxels and fit it with a truncated normal distribution, which we use to sample random $E(B-V)$ values for use in our mock training dataset. While this approach does not associate an $E(B-V)$ value with low-S/N spaxels, these regions are likely to fall towards the range of extinction values already sampled by our distribution.

Due to the low peculiar velocity of M33, the \oi$\lambda6300$ line is not sufficiently redshifted from its airglow counterpart and is therefore strongly affected by sky subtraction residuals and generally not recoverable in our MUSE data. We therefore removed it from the mock spectra to avoid the network interpreting airglow emission as \oi$\lambda6300$ from the target.

The architecture of the network used in this case is akin to the one already described in the multi-label case (Sect. \ref{sec:mlspec}), with an increase of the dropout rate to $30\%$ to help with the model's generalisation.
Since this network is trained with a multi-label type of dataset, we evaluate its performance as in Sec. \ref{sec:mlspec}. We find that the average median error on the prediction of each class' contribution for the mock test dataset is $\sim9\%$. We also measure an f1-score of 0.80 considering only the test instances with the prevalent class contributing to at least 50$\%$ of the spectrum and comparing that class with the dominant one predicted by the network. This performance is comparable to that of the previously described low S/N and the multi-label cases (Sec. \ref{sec:test}). 

Before testing the model on our data, we perform two additional pre-processing steps.
Missing data (NaNs) in the spectra are set to zero, and the input spectra are standardised (the mean is subtracted and then they are divided by the standard deviation) before input into the network. This is the same standardisation process performed on the mock spectra and is a fundamental step because these spectra may have a non-negligible contribution from the
continuum. However, this contribution is sufficiently small with respect to the emission lines, that standardising the spectrum makes a detailed continuum subtraction unnecessary.
In total, it takes $\sim20$ hours to complete the classification for every spaxel in the datacube ($2,356,584$ spectra) on a single GPU.

\subsection{Classification of the IFS datacube} 
\label{sec:ifsclassif}

\begin{figure*}[p]
    \centering
    \includegraphics[width=0.90\linewidth]{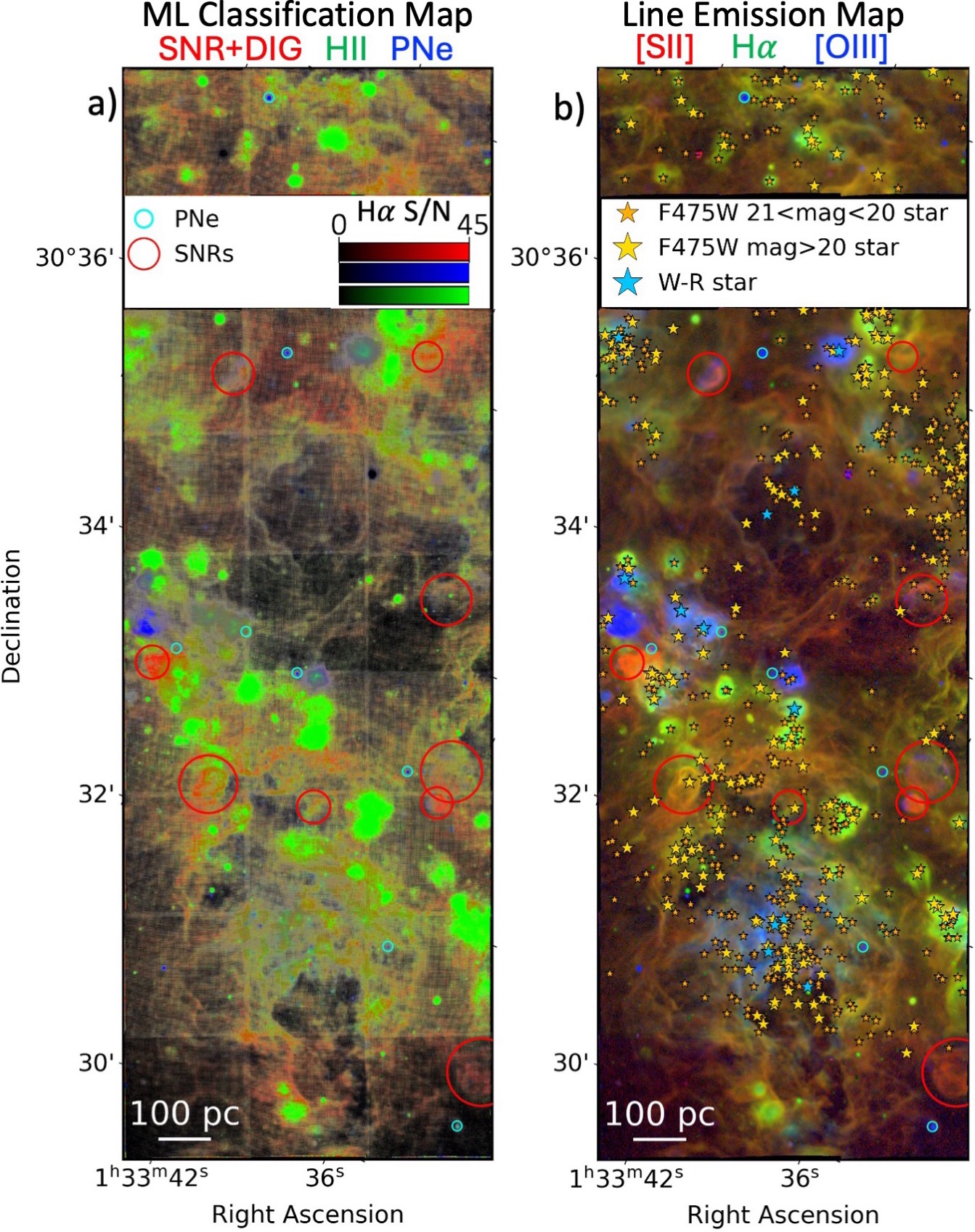}
    \caption{Comparison between the neural network classification of the single-spaxel spectra and a map of three key emission lines. On the left side (panel a), we show our neural network's prediction map for the observed field of M33. Each colour quantitatively represents the predicted contribution to each pixel's spectra from an \hii region (green), a PN (blue), and a SNR plus DIG (red). We overlay a S/N map (in shades of grey), derived from the \ha\ flux map and its associated error map.
    On the right side (panel b), we show an RGB map of three key emission lines for the entire observed field of M33, with the red scale representing the \siit flux, the blue the \oiii$\lambda5007$ flux, and the green the \ha\ flux. The gold star symbols mark massive stars brighter than 20 mag in the F475W Hubble Space Telescope (HST) filter and the smaller orange stars those with magnitude between $21$ and $20$ in the same filter from \cite{Williams2021}. The light blue star symbols mark Wolf-Rayet stars \citep{Neugent2011}.
    In both images, the red circles mark SNRs from the \cite{Lee2014} and \cite{Long2018} catalogues, while the smaller teal circles show the PNe selected by \cite{Ciardullo2004}.
    }
    \label{fig:M33ML_images_mix}
\end{figure*}

We applied the `single spaxel' model to the entire IFS datacube and show the result of the classification process in Fig. \ref{fig:M33ML_images_mix}. In particular, the figure shows the predicted classification for each spaxel of the observed field of the M33 galaxy (panel a) compared to the map of the emission from three key lines, \ha, \siit, and \oiii$\lambda5007$ (panel b).
The prediction map is a three-colour rgb image, where the green colour scale is associated with the \hii regions class, the blue one with PNe, and the red one with SNRs and DIG classes summed together. 
Each pixel in the image is coloured based on the neural network's predicted probability for the spectrum within the corresponding spaxel, with the colour increasingly aligning with the representative colour of each class as its predicted contribution increases.
The motivation for considering the SNR and DIG probabilities together is further discussed later in this section and is mostly driven by the network's difficulty in clearly distinguishing between these two classes.

In Fig. \ref{fig:M33ML_images_mix}a, we also overlap a S/N map in grey scale for a better interpretation of the prediction map. 
Darker areas on the map indicate lower S/N levels. The square grid which is visible in the S/N map represents the edges of the MUSE pointings and becomes apparent because the overlap regions between pointing have intrinsically higher S/N because of their deeper integrations.

By comparing panels a and b in Fig. \ref{fig:M33ML_images_mix}, we conclude that the network identifies \hii regions across the entire field. In particular, regions classified as \hii\ regions by the network correspond well to \ha-bright regions (which appear green in Fig. \ref{fig:M33ML_images_mix}b) hosting ionising stars (Feltre et al. in prep., \citealt{Williams2021}), which are shown with a star symbol in Fig. \ref{fig:M33ML_images_mix}b. 

\begin{figure}
    \centering
    \includegraphics[width=1\linewidth]{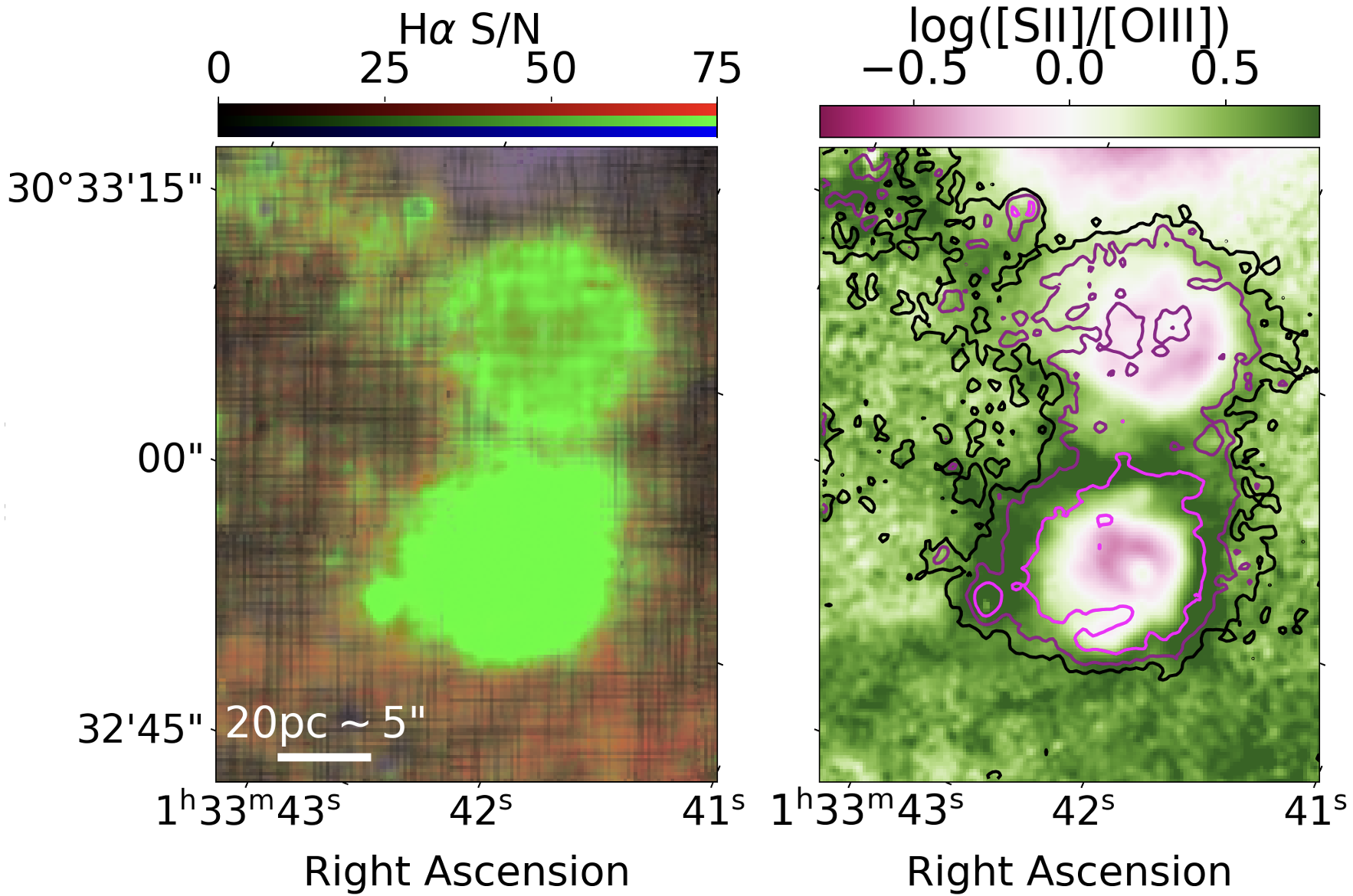}
    \caption{Zoom-in of a region of interest in the M33 field, focusing on two \hii regions. In the left classification map, we quantitatively represent the predicted contribution to each pixel's spectra from an \hii region in green, a PN in blue, and a SNR/DIG in red. We overlay a S/N map (in shades of grey), derived from the \ha\ flux map and its associated error map. The right panel shows a \siit/\oiii$\lambda5007$ ratio map. The black, purple, and magenta contours in this panel enclose a model-predicted \hii region probability of 0.4, 0.6, and 0.9, respectively.}
    \label{fig:3hiiregions}
\end{figure}

To investigate where the model sets the \hii region boundaries we examine two prototypical \hii regions shown in Fig. \ref{fig:3hiiregions} which are among the most luminous (4.4 and 9 $ \times 10^{17}  \mathrm{erg~s^{-1}}$, top and bottom, respectively) identified in our MUSE data. They show a nearly spherically symmetric geometry, with emission from higher/lower ionisation lines decreasing/increasing from the centre outwards, and exhibit clear ionisation fronts, as traced by the \sii/\oiii\ ratio (Feltre et al. in prep.). More details on the spectral properties of these regions are provided in \cite{Pellegrini2012} and Feltre et al. (in prep.). Here we compare the neural network classification with the \sii/\oiii\ line ratio map. The black, purple, and magenta contours in this map highlight where the model predicts a probability of \hii region equal to $40\%$, $60\%$, and $90\%$, respectively.
The comparison shows that the predicted \hii regions extend up to the ionising front traced by the high \sii/\oiii\ ratio. Thus, our \hii\ region classification successfully classified the whole nebula, out to the ionisation front, when considering a probability of 0.6.

\begin{figure*}
    \centering
    \includegraphics[width=0.98\linewidth]{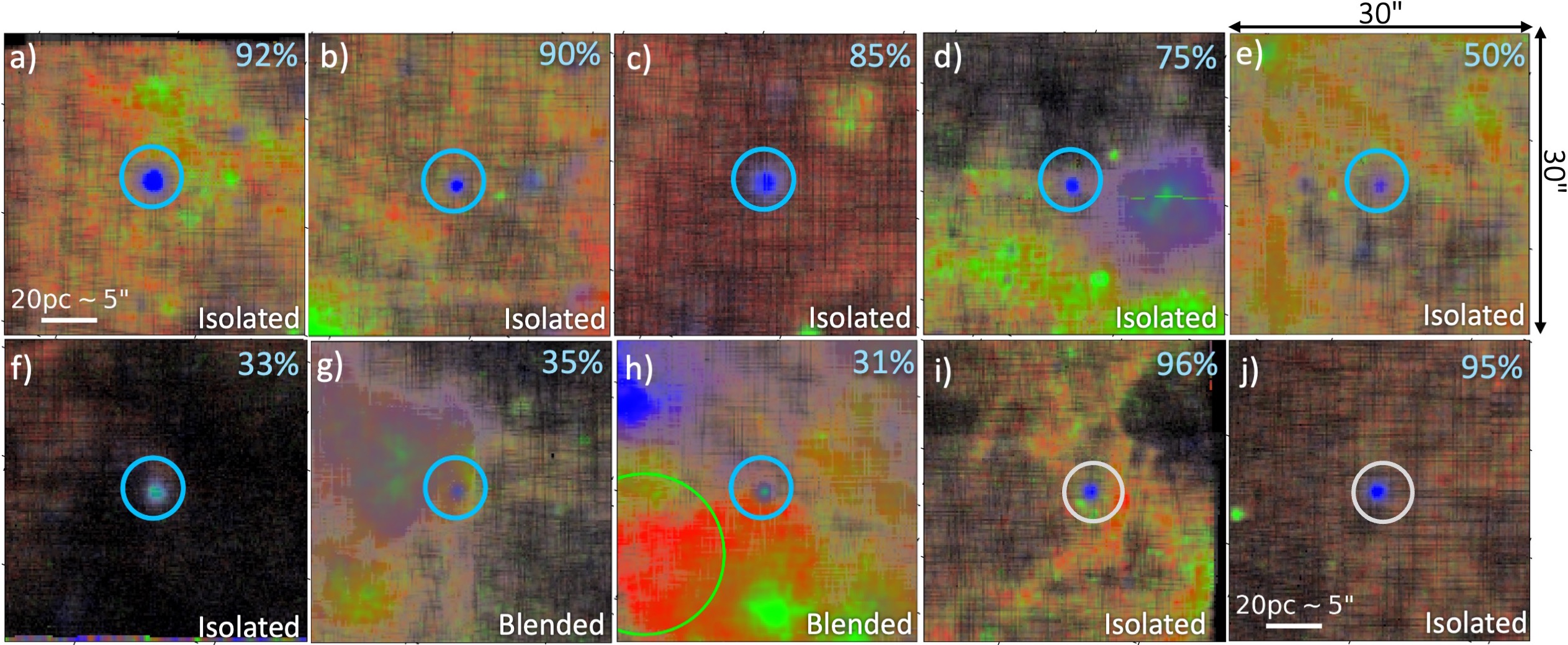}
    \caption{Cutouts of the classification map focusing on the \cite{Ciardullo2004} PNe, identified by the light blue circles (panels a-h) and two additionally identified PNe, marked in grey (panels i,j). Each colour of the RBG map quantitatively represents the predicted contribution to each pixel's spectra from an \hii region (G), a PN (B), and a SNR/DIG (R). We overlay a S/N map (in shades of grey), derived from the \ha\ flux map and its associated error map.
    On the top right of each cutout, we write the PN contribution for the identified PNe averaged over an aperture of $1.5"$ of diameter centred on the probability peak. On the bottom left, we specify whether the PN is isolated or its emission appears to be blended to that of other regions.
    The green circle in panel h marks a nearby SNR from the \cite{Lee2014} and \cite{Long2018} catalogues.}
    \label{fig:pne_cutouts}
\end{figure*}

We then test the ability of the network to detect the PNe previously identified in \cite{Ciardullo2004}. Firstly, we consider the \quotes{isolated} PNe, whose classification maps are visible in panels a-f of Fig. \ref{fig:pne_cutouts}.
We find that our model predicts a strong PN probability ($\gtrsim75\%$) for four of these PNe and a $50\%$ probability of another one (i.e. PN is still the dominant class). Only the PN in panel f has another class (\hii region) as the prevalent one ($\sim50\%$).
Additionally, where PNe occur in the vicinity of another bright region, and therefore some overlap along the line-of-sight seems probable, the network predicts a more mixed classification and a lower PN probability ($\sim30\%$). This is the case for the PNe in panels g and h of Fig. \ref{fig:pne_cutouts}. For these, the prevalent class is \hii region (40-50$\%$ probability).
Furthermore, the network detects new PNe that were missed in the literature catalogues with very high probabilities ($\gtrsim90\%$). Two examples are visible in panels i and j of Fig. \ref{fig:pne_cutouts}.

Aside from the confidently identified \hii regions and PNe, the remainder of the field's spectra are mainly classified as SNRs or DIG.
However, we observe that the network struggles with distinguishing SNR from DIG spectra. 
Indeed, as shown in the confusion matrices in Fig. \ref{fig:cm_all}, we have already observed that the network exhibits some degree of confusion between SNRs and DIG spectra.
Additionally, in this specific case, the \oi$\lambda6300$ line is not considered in our model because of the overlap with airglow emission (Sec. \ref{sec:tailoring}). Both traditional diagnostics \citep{Kopsacheili2020} and our neural network (as discussed in Sec. \ref{sec:actmax}) identify \oi$\lambda6300$ as a critical emission line for the classification of SNRs. In its absence, the network is compelled to rely predominantly on the \siit doublet, which proves insufficient to discriminate between the two classes.
This limitation arises because the emission line ratios in the DIG are almost always consistent with low-ionisation emission regions (LIER), but also shock models \citep{Allen2008} significantly overlap with the LIER region of the BPT diagram, as observed in our case (see Fig. \ref{fig:bptplots}).
These considerations motivate our decision to sum the prediction probabilities for SNR and DIG when discussing our results. 

Considering this, our model correctly classifies the SNRs documented in the literature (red circles in  Fig. \ref{fig:M33ML_images_mix}, from \citealt{Lee2014,Long2018}) as belonging to the SNR+DIG class with varying degrees of confidence. Similarly, the spectra from part of the filamentary structure visible at low S/N in the line emission map are classified as \hii region mixed with the SNR$+$DIG class.

\begin{figure}
    \centering
    \includegraphics[width=0.9\linewidth]{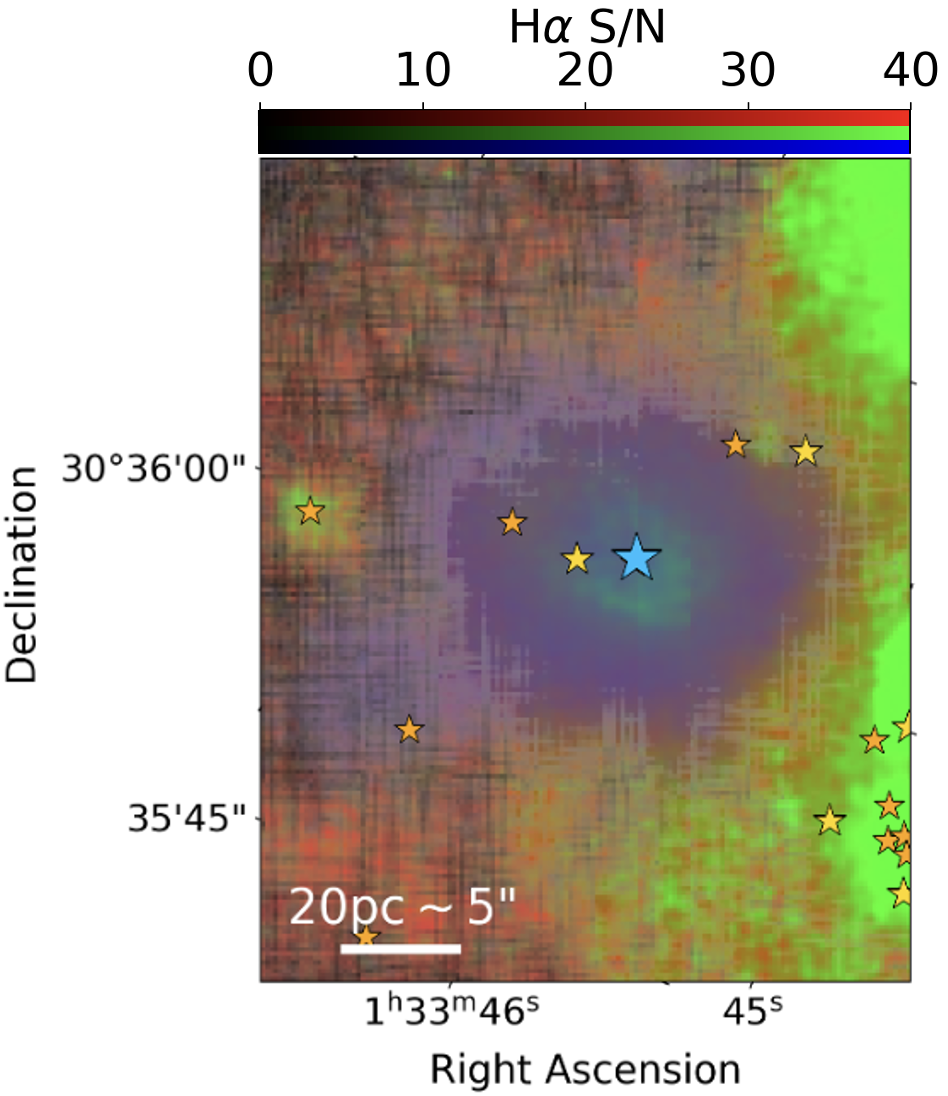}
    \caption{Cutout of the classification map focusing on an \oiii$\lambda5007$ bright \hii region hosting a Wolf-Rayet star \citep{Neugent2011}, marked by the light blue star symbol. The colours in the map quantitatively represent the predicted contribution to each pixel's spectra from an \hii region (green scale), a PN (blue), and a SNR/DIG (red). We overlay a S/N map (in shades of grey). The additional star symbols in yellow and gold represent massive stars respectively brighter than 20 mag and between $21$ and $20$ mag in the F475W HST filter \citep{Williams2021}.}
    \label{fig:oiiibrighthii}
\end{figure}

We also observe some bright regions with a mix of contributions from both PNe and \hii\ region (as in Fig \ref{fig:oiiibrighthii}). 
Despite the fact that PN may be the most probable classification, these regions have an extended morphology and are clearly ionised by either Wolf-Rayet (WR) stars \citep{Neugent2011} or hot young stars \citep{Williams2021}.
Indeed, the bright ultraviolet (UV) flux from WR stars (among the hottest and most luminous stars in the local Universe) creates a highly ionised \hii region, or \quotes{WR nebula} (e.g. \citealt{Chu1981, Tuquet2024}). The WR star's hard spectrum results in strong emission lines from high-ionisation species such as \oiii, which is also a prominent feature in PNe. Because of their underlying spectral similarity, WR nebulae are often classified as PNe by our neural network.

Finally, we conclude that the neural network is good at distinguishing \hii regions and PN spectra that show brighter high-ionisation line emission, from SNRs and DIG spectra, which are characterised by low-ionisation line emission.
We compare these classifications with traditional diagnostics in Sec. \ref{sec:ifsdiscussion}.

\section{Discussion} 
\label{sec:discussion}

In this section, we explore the interpretation of our neural network approach through activation maximisation to identify the spectral features used for classification. We also discuss the limitations of our neural network model, in particular the misclassifications arising from photoionisation and shock model mismatches, and propose future work on domain adaptation to address these challenges.

\subsection{Interpreting the model with activation maximisation} \label{sec:actmax}

\begin{figure*}[p]
    \centering
    \begin{minipage}{\linewidth}
        \centering
        \includegraphics[width=0.95\linewidth]{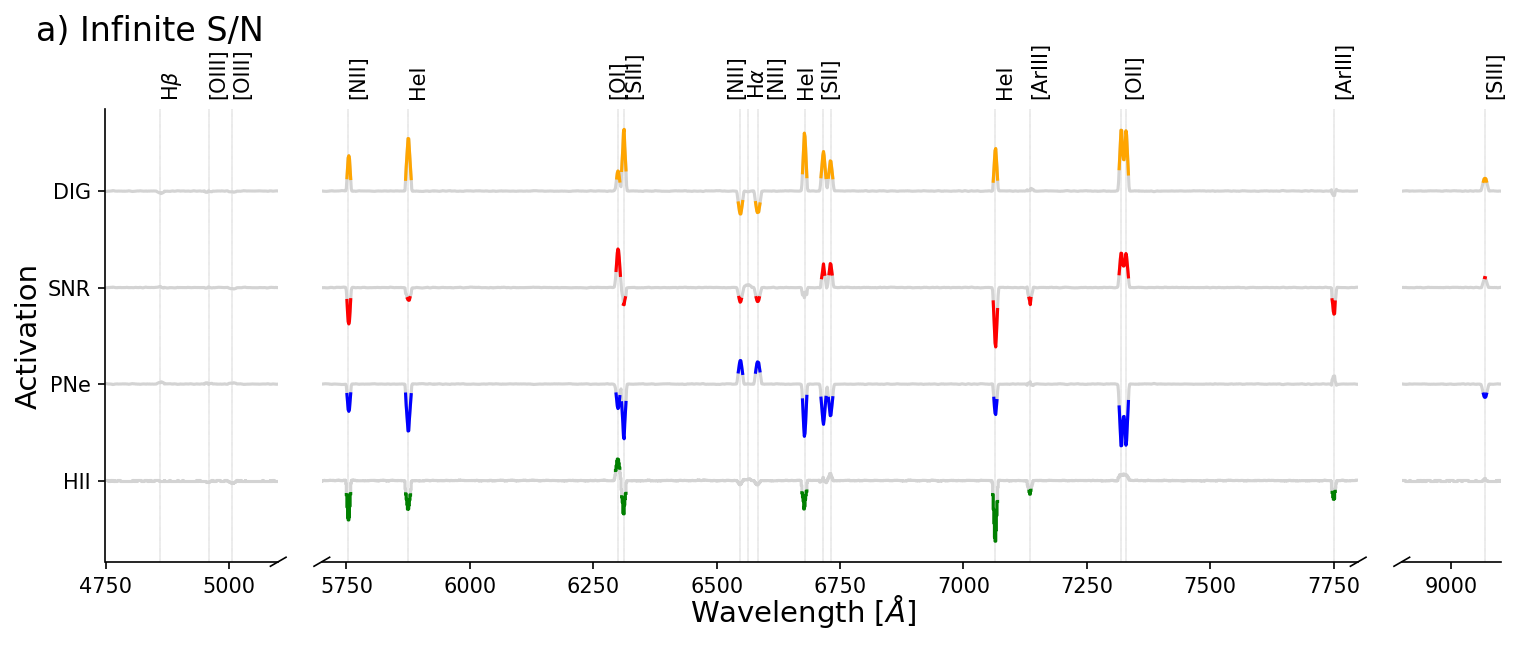}
    \end{minipage}
    
    
    \begin{minipage}{\linewidth}
        \centering
        \includegraphics[width=0.95\linewidth]{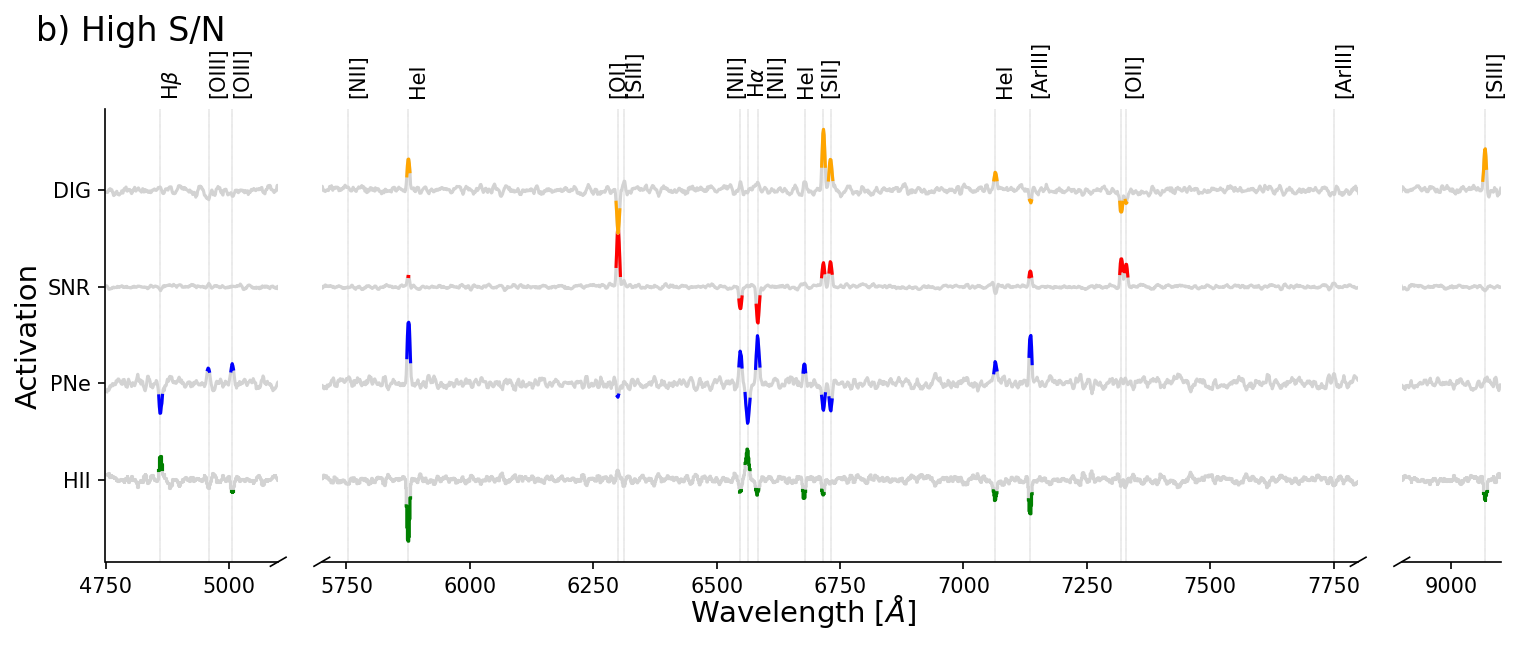}
    \end{minipage}
    
    
    \begin{minipage}{\linewidth}
        \centering
        \includegraphics[width=0.95\linewidth]{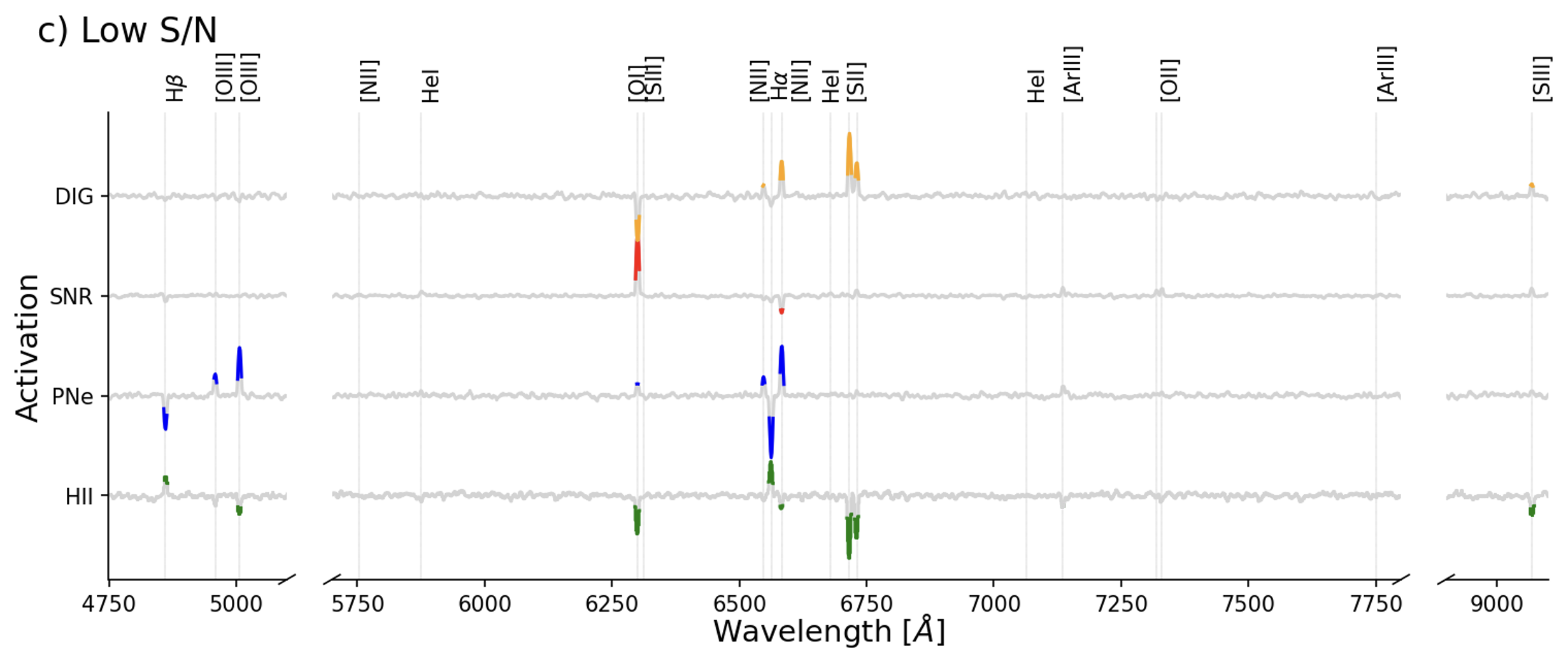}
    \end{minipage}
    
    \caption{Activation maximisation maps corresponding to three different mock datasets with different S/N for the four different classes of nebulae. For ease of visualisation, we show in colour the spectral channels where the absolute value of the activation is larger than $15\%$ of the maximum value of the activation spectrum. The key emission lines are labelled. Panel a): activation map for the network trained with mock spectra with infinite S/N. Panel b): The high S/N mock spectra case. Panel c): the low S/N spectra case.}
    \label{fig:actmaps}
\end{figure*}

Activation maximisation is a method used to visualise and interpret the features in the data that specific neurons in a neural network have learned to recognise. By adjusting the input spectrum to enhance the activations of the target neurons, activation maximisation produces synthetic spectra that are optimised to trigger strong responses from those neurons. The generated spectra provide visual insights into the features that the resulting classification is based on.

We show activation maximisation maps (i.e. spectra) for three of our mock single-label datasets (infinite S/N, high S/N, low S/N) in Fig. \ref{fig:actmaps}. These ideal spectra show both positive and negative peaks located at the position of some of the emission lines present in the training dataset. Regardless of the sign of the values, the more prominent the peak, the larger influence that line will have on the network's classification. In all cases, the network correctly neglects the channels devoid of emission lines.

When trained with the infinite S/N dataset, the network bases its classification on faint emission features, including the auroral lines of the main metal ions (\nii$\lambda5755$, \siii$\lambda6321$, \oii$\lambda7320,30$) and the \hei recombination lines at $\lambda6678$, $\lambda7065$.
Thus, according to our model, these faint lines are the most suitable to efficiently classify emission-line regions.

In the high S/N case, the network starts to shift its focus towards the brighter lines in the spectrum for all four classes.
Indeed Fig. \ref{fig:actmaps}, panel b, shows that the lines the neural network uses to classify PNe, for example, are both some of the faint lines (such as \hei$\lambda5876$, \ariii$\lambda7135$, \hei$\lambda7065$), and some of the bright ones (such as \ha, \oiii$\lambda\lambda4959,5007$, \nii$\lambda6548,84$ and \hb).

In the low S/N case, strong emission lines such as \ha, \hb, \oiii$\lambda4959, 5007$, \nii$\lambda6584$ become even more emphasised, while the fainter lines lose relevance (Fig. \ref{fig:actmaps}, panel c). Indeed, the network is now relying exclusively on \ha, \hb, \nii$\lambda6548,84$, \oi$\lambda6300$, \siit, and the \siii$\lambda9069$ lines to classify the mock spectra.
In this last case, we find that the classification strategy learnt by the network tends to mirror that of the traditional techniques. 
For example, the characteristic features the network positively reacts to in \hii regions spectra are the Balmer recombination lines, while it characterises DIG spectra by the lower ionisation lines of \nii\ and \sii. PNe are identified using \oiii, \ha\ and \nii, as suggested by \cite{Ciardullo2002}, and SNR by \oi/\ha\ as suggested by \cite{Kopsacheili2020}. 
The activation maps of the network trained on the high S/N dataset with extinction exhibit intermediate characteristics between those of the models trained on the high S/N and low S/N datasets. 

These findings highlight important advantages of our neural-network-based approach over traditional diagnostic diagrams. 
While the network relies on spectral features similar to those employed in classical techniques, our method enables the emergence of new diagnostic trends in high-S/N regions.  
Unlike diagnostic diagrams that rely on pre-computed flux ratios in low-dimensional (2D or 3D) spaces, our approach directly ingests spectra. This allows the model to exploit the full information content of the spectrum, naturally exploring a multi-dimensional diagnostic space with potentially non-linear relationships between features. Consequently, the network can uncover subtle, complex patterns in the data that traditional approaches might overlook.

\begin{figure*}[ht]
    \centering
    \includegraphics[width=0.95\linewidth]{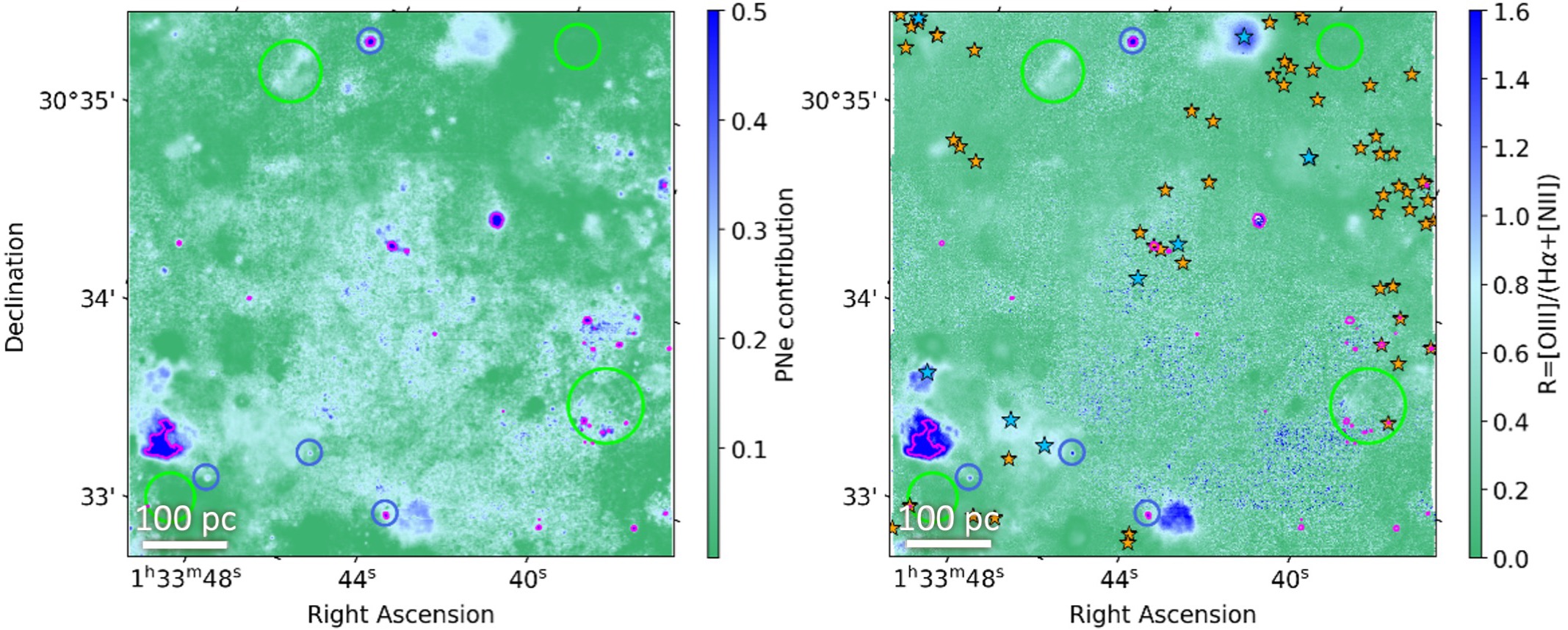}
    \caption{Comparison between the PNe neural network classification and the one with a traditional method from the literature for a section of the observed M33 field. On the left is the PN contribution map as predicted by the network. Colours tending to blue show a higher probability of PN, while those tending to green show a lower one.
    On the right, a map coloured based on the $R = \rm [OIII]\lambda5007/(H\alpha + [NII]\lambda6584)$ ratio from \cite{Ciardullo2002}, where all the pixels coloured in blue are predicted to be PNe. The yellow star symbols mark O type ionising stars from \cite{Williams2021}, selected based on flux and colour. Instead, the light blue ones mark Wolf-Rayet stars \citep{Neugent2011}.
    In both images, the blue circles show the \cite{Ciardullo2004} PNe, while the green circles mark the \cite{Lee2014} and \cite{Long2018} SNRs. The pink contours mark the points with a predicted PN contribution of 0.5.}
    \label{fig:pne_compare}
\end{figure*}

\begin{figure*}[ht]
    \centering
    \includegraphics[width=0.95\linewidth]{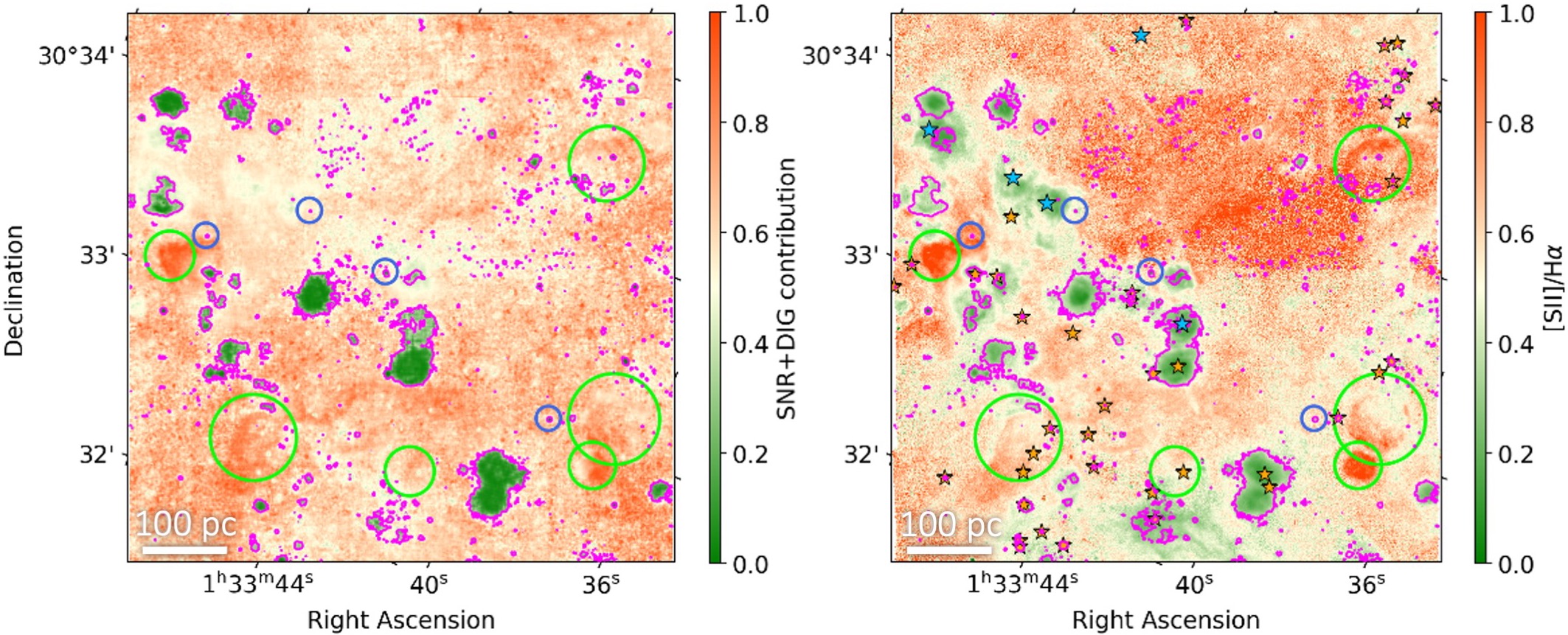}
    \caption{Comparison between the SNR+DIG neural network classification and the one with a traditional method from the literature for a section of the observed M33 field. On the left is the SNR+DIG contribution map as predicted by the network. Colours tending to red show a higher probability of SNR or DIG, while those tending to green a lower one.
    On the right, a map coloured based on the \siit/\ha\ ratio. A higher value of the ratio suggests brighter \sii\ emission with respect to \ha. The yellow star symbols mark O type ionising stars from \cite{Williams2021}, selected based on flux and colour. Instead, the light blue ones mark Wolf-Rayet stars \citep{Neugent2011}.
    In both images, the blue circles show the \cite{Ciardullo2004} PNe, while the green circles mark the \cite{Lee2014} and \cite{Long2018} SNRs. The pink contours mark the points with a predicted SNR+DIG contribution of 0.6.}
    \label{fig:shk_compare}
\end{figure*}

\subsection{Evaluating the IFS data classification against traditional diagnostics} 
\label{sec:ifsdiscussion}

We further assess the IFS datacube classification by comparing it with the one performed with traditional methods.
In particular, we consider two traditional diagnostic line ratios: $R=\text{\oiii}\lambda5007/(\text{H}\alpha+\text{\nii}\lambda6548)$  for PNe \citep{Ciardullo2002}, and \sii$\lambda\lambda6717,6730/$H$\alpha$ for SNRs and DIG \citep{DOdorico1980}. 

Fig. \ref{fig:pne_compare} shows the PNe probability map generated by the network alongside a map of the $R$ value for a section of the M33 field.
In the probability map, blue pixels indicate regions with a predicted probability of being PNe higher than $50\%$.
\cite{Ciardullo2002} consider PNe all the nebulae with $R>1.6$, which are coloured in blue in the right panel. The magenta contours on both maps highlight areas with a predicted PNe contribution of $50\%$, providing a direct visual comparison between the network predictions and the $R$-based classification.
The two panels of Fig. \ref{fig:pne_compare} show how well the areas with a strong likelihood of being PNe trace the areas with higher R values, considering both PNe and \oiii$\lambda5007$ bright \hii regions. In particular, this threshold on PNe probability traces most of the PNe in \cite{Ciardullo2004}, with the exception of those embedded in more complex regions (those of panels e-g of Fig. \ref{fig:pne_cutouts}).
For a more quantitative comparison, we used the intersection over union (IoU) metric, which measures the overlap between two regions by dividing the area of their intersection by the area of their union (0 indicates no overlap, 1 perfect alignment). We do this considering the area of all the spaxels in the entire field with a PNe probability $>0.5$ and that of all the spaxels with $R>1.6$, obtaining a IoU value of $\sim0.69$

Similarly, in Fig.\ref{fig:shk_compare} we study the SNR+DIG region probability map together with a map of the \sii/\ha\ ratio in another section of the M33 MUSE field. We consider the fact that a \sii/\ha\ ratio closer to one corresponds to regions that are more likely to be shock ionised \citep{Mathewson1972, DOdorico1980}. Again, there is a satisfactory conformity between the two maps, especially considering (some) of the regions in the green circles which are SNR from the literature.
We again measure the IoU for the entire field, considering the SNR+DIG probability threshold at 0.6 (so that the \hii regions and PNe are clearly separated from the SNR/DIG) and the \sii/\ha\ ratio threshold at 0.5, retrieving a IoU value of $\sim0.73$, confirming a good agreement between the two maps.

\subsection{Limitations and future work}

The findings presented here are subject to certain limitations that warrant further investigation.
Firstly, all the used emission-line region models are single-cloud models, that is, they describe the integrated spectrum of a single nebula \citep{Perez-Montero2014,Delgado-Inglada2014}. On the other hand, we compare them to spatially resolved spectra of these regions, which show complex structures within the observed nebulae (Feltre et al. in prep.), with the exception of PNe, which are unresolved. Therefore, the assumptions of single-cloud models (e.g. uniform density, spherical geometry) are not realistic in most cases and are not sufficient to interpret observations of spatially resolved nebulae.
To address this limitation, it would be necessary to train our neural networks with photoionisation and shock models that better describe the elaborate structure of these nebulae, such as three-dimensional models \citep{Ercolano2005,Ercolano2007,Jin2022} or state-of-the-art multi-cloud models (e.g. HOMERUN, \citealt{Marconi2024}). 
Nevertheless, despite these simplifications, single-cloud models still provide a meaningful classification, even if they do not capture the full complexity of spatially resolved structures. In Appendix \ref{sec:AppB}, we discuss the application of our ML method to integrated data.

Additionally, misclassifications can arise from a discrepancy between the photoionisation and shock models and observed nebulae. We have verified that this discrepancy is not large by comparing $\sim2000$ real spectra extracted randomly from the cube from areas with S/N$\geq7$ on the \ha\ line and finding the model for which the sum of the absolute value of the residuals is minimum. The reduced $\chi^2$ of these fits is always below 2, with average equal to 0.5, ensuring that there is at least one mock spectrum accurately reproducing a real one within twice the noise levels of the observed spectra.

To address the problem of model mismatch within an ML framework,  \cite{Belfiore2024} implemented domain adaptation techniques \citep{pmlr-v37-ganin15}. Domain adaptation bridges the gap between photoionisation and shock
models and observed ionised nebulae by learning representations valid across both domains. Implementing such techniques can significantly improve the robustness of neural network models, reducing the biases introduced when training only on theoretical predictions.
This can be achieved with the use of a DANN (domain-adversarial neural network, \citealt{ganin2016domain}). The goal of a DANN is to maximise the performance of a model trained on labelled data in the source domain (our synthetic spectra) on an unlabelled target domain (our observed spectra) by extracting domain-invariant features (those shared by both source and target domain).

\begin{figure}
    \centering
    \includegraphics[width=\linewidth]{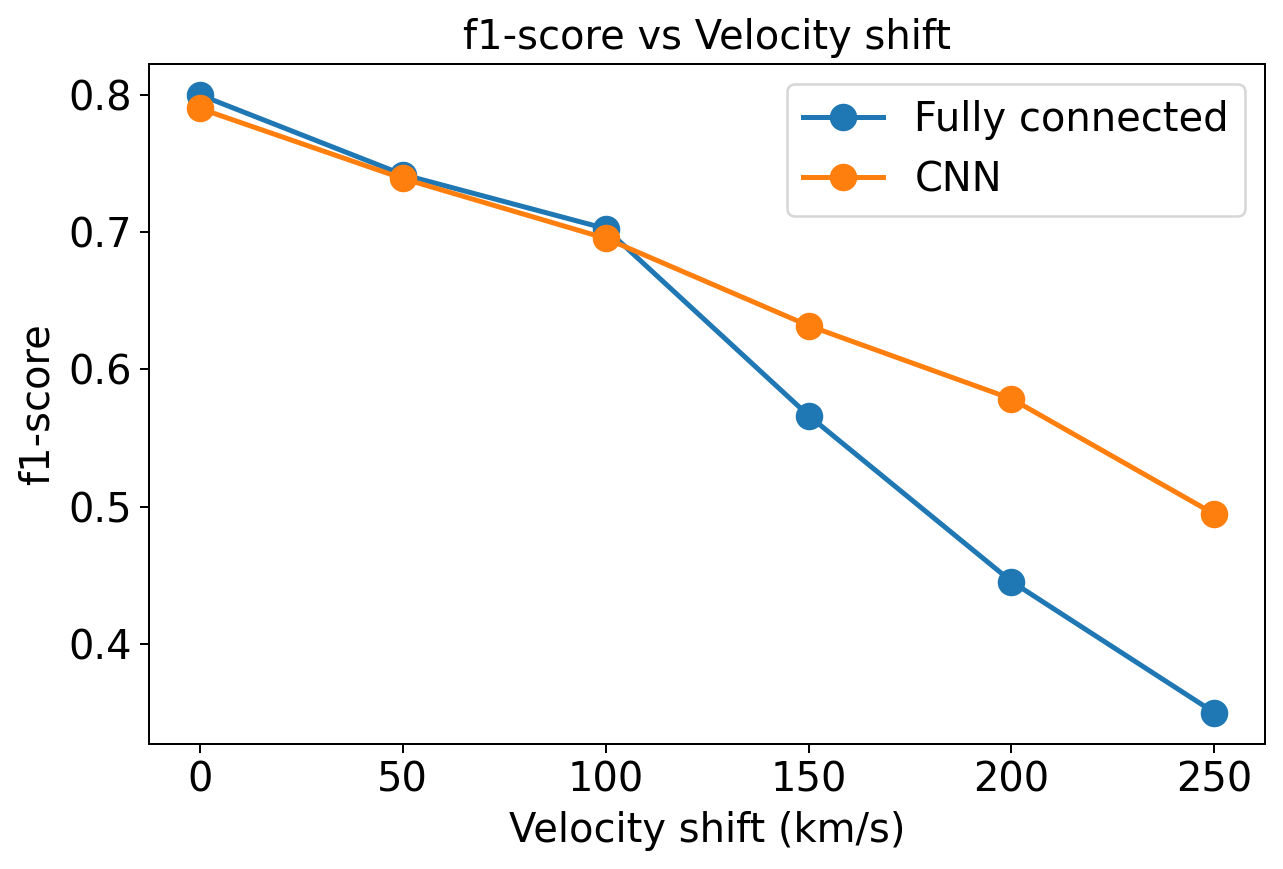}
    \caption{Evolution of the f1-score for different velocity shifts in the mock test dataset for the single spaxel fully connected model and a CNN trained on the same dataset.}
    \label{fig:velshifts}
\end{figure}

Finally, we consider the resilience of the \quotes{single spaxel} model to line shifts in the dataset since a fully connected neural network is not inherently shift invariant. To do so, we measure the f1-score as in Sect.\ref{sec:mlspec} on multi-label mock test spectra with different velocity shifts ($\pm50,100,150,200,250$ km/s).
For comparison, we apply the same test with a CNN trained on the same tailored mock dataset (Sect.\ref{sec:tailoring}). The CNN is composed of two convolutional layers (one with with five kernels of dimension equal to two wavelength channels and one with 50 kernels of dimension equal to ten channels) both followed by a max pooling layer with pooling size equal to four, a flattening layer, two dense layers composed of eight units each followed by a 10\% dropout layer, and then the output layer.
The results of this evaluation are presented in Fig.\ref{fig:velshifts}.
The fully connected model maintains a strong performance (f1-score $\sim70\%$) up to $\sim100$ km/s (equivalent to a shift of approximately two wavelength channels in our MUSE dataset), which corresponds to the maximum velocity shift observed in our M33 field. 
In contrast, due to its inherent translational invariance, the CNN outperforms the fully connected at larger shifts.
Thus, for our dataset, we find the fully connected network to be sufficiently robust for the classification. However, a CNN should be considered for broader applications.
Since this test is carried out at the inference stage, a better performance could be achieved by adding velocity shifts in the training dataset.

\section{Summary and conclusions}
\label{sec:conclusions}

The study of emission-line regions in extragalactic environments is critical for understanding galaxy evolution.
We demonstrate the potential of using artificial neural networks to classify nebular regions by their spectra, overcoming some of the issues of traditional diagnostics. These limitations include rigid classification boundaries, the challenge of classifying spectra with low signal-to-noise ratios and intermediate properties (e.g. due to superposition), the limited use of the information contained in the data, and the time-consuming step of spectral line fittings.

We train and test our neural network with mock datasets created starting from emission-line intensities in photoionisation and shock models, emulating instrumental and physical observational effects, and simulating a range of noise and extinction levels. 
We also develop a multi-label dataset to train the network in recognising cases where the spectra may display features that overlap between different classes (e.g. caused by superposition along the line-of-sight).
We evaluate the network's performance across the different training datasets and use activation maximisation maps to investigate the features in the spectra on which the network relies to carry out its classification.
Our main findings after the training and testing of the network are:
   \begin{itemize}
      \item With high S/N spectra (S/N = 100) the network shows excellent classification performance (f1=0.97) and relies on faint spectral lines, such as \nii$\lambda5755$, \hei$\lambda5876$, \sii$\lambda6312$, \hei$\lambda7065$, \oii$\lambda7320$, \oii$\lambda7330$, and \ariii$\lambda7751$.
      \item When noise and extinction are introduced, the network relies more on brighter lines to maintain classification accuracy. This approach resembles traditional techniques, as it draws on lines typically used in single- and multi-ratio diagnostic methods. This suggests that the model has, to some extent, learned features of traditional classification approaches \citep{DOdorico1980, Baldwin1981, Ciardullo2002, Kopsacheili2020}.
   \end{itemize}

We also run the network on real spectra extracted from a MUSE observation of a field of the nearby galaxy M33. 
We classify over two million single-pixel spectra.
The key results from the tests on real data are:
\begin{itemize}
    \item The network is able to effectively distinguish \hii regions and PNe spectra from SNRs and DIG ones across the entire observed field, confidently identifying most of the PNe from the literature \citep{Ciardullo2004}.
    \item The difficulty in distinguishing between SNRs and DIG spectra possibly arises from the missing \oi$\lambda6300$ line, which according to both our test and the literature \citep{Kopsacheili2020} is key in this discrimination. Taking this into account, we also retrieve the literature SNRs \citep{Lee2014,Long2018}.
    \item The comparison of the spaxel-wise classification results with commonly used diagnostic line ratio maps \citep{Ciardullo2002,DOdorico1980} further supports the network’s alignment with traditional methods. 
\end{itemize}

In conclusion, our approach demonstrates that neural networks can effectively classify emission-line regions by dynamically selecting different sets of spectral lines based on the available data quality. 
Additionally, by leveraging the three-dimensional nature of IFS data, our method allows for the definition of physically motivated spatial boundaries for the nebular regions and accounts for the effects of projection and line-of-sight superposition. These capabilities represent a significant advancement over traditional classification methods. More broadly, this work highlights the potential of ML as an invaluable tool for the analysis of large IFS data, paving the way for broader applications in the automated analysis of complex astrophysical datasets.

\begin{acknowledgements}
Based on observations obtained at the ESO/VLT Programme ID 109.22XS.001 (PI:Cresci).
FB acknowledges support from the INAF-Fundamental Astrophysics programme 2022 \& 2023.
AF acknowledges the support from project "VLT- MOONS" CRAM 1.05.03.07, INAF Large Grant 2022 "The metal circle: a new sharp view of the baryon cycle up to Cosmic Dawn with the latest generation IFU facilities" and INAF Large Grant 2022 "Dual and binary SMBH in the multi-messenger era”.
EB and GC acknowledges the support of the INAF Large Grant 2022 ""The metal circle: a new sharp view of the baryon
cycle up to Cosmic Dawn with the latest generation IFU facilities"
The work of {AB} was funded by Progetto ICSC - Spoke 2 - Codice CN00000013 - CUP I53C21000340006 - Missione 4 Istruzione e ricerca - Componente 2 Dalla ricerca all'impresa – Investimento 1.4.
GT acknowledges financial support from the European Research Council (ERC) Advanced Grant under the European Union’s Horizon Europe research and innovation programme (grant agreement AdG GALPHYS, No. 101055023).
GV acknowledges support by European Union’s HE ERC Starting Grant No. 101040227 – WINGS.

\end{acknowledgements}


\bibliographystyle{aa} 
\bibliography{aa54196-25} 

\begin{appendix}

\section{Selection of photoionisation and shock models}
\label{sec:AppA}
To model the line fluxes of \hii regions, we used the \texttt{HII\_CHIm} model grid from \cite{Perez-Montero2014}. This is a small grid of \hii\ region models, with input ionising spectra from POPSTAR models \citep{Molla2009} of a single instantaneous burst of star formation with an age of 1 Myr. The selected parameters and their values are shown in Table \ref{tab:models}, leading to a total of 452 models.

For PNe, we adopted the \texttt{PNe\_2014} grid \citep{Delgado-Inglada2014}, which covers a wide range of physical parameters and is representative of most of the observed PNe. The chosen input parameters are reported in Table \ref{tab:models}, leading to a selection of 216 PNe models.

Models for SNRs were taken from the 3MdBs (3MdB shock) table \citep{Alarie2019}. These models are not specifically designed to represent SNRs but rather radiative shocks in general. 
This may affect the classifications of real data, as shocks may also occur elsewhere in the ISM.
We adopted the models from the \cite{Allen2008} grid using the parameters in Table \ref{tab:models}, leading to a total number of 297 shock models.

Lastly, DIG models were recovered from the \texttt{DIG\_HR} grid \citep{Flores-Fajardo2011} in 3MdB. The project consists of a grid of plane-parallel, ionisation-bounded models in which two ionisation sources are combined: O-B stars and hot low-mass evolved stars (HOLMES). The SED representing the O-B stars is taken from the evolutionary stellar population synthesis code \texttt{STARBURST99} \citep{Leitherer1999}, while the SED representing HOLMES is from \texttt{PEGASE.2} \citep{Fioc1997}. 
A sub-sample of the available models representing standard DIG conditions was selected as described in Table \ref{tab:models}, leading to the selection of 3928 DIG models.

Our selection criteria generated a small and unbalanced set of models for the different classes, with gaps between the models in the parameter space. For each class, we therefore generated new models using a linear combination of two models from the same class, with the combination weights chosen randomly. The resulting ensemble of interpolated models is capable of populating the observational parameter space more densely. 
The final set of models consisted of $35\,000$ models for each of the four types of emission-line regions. 

\begin{table}[tb]
    \centering
    \caption{Parameter selection for the photoionisation and shock models.}
    \resizebox{\linewidth}{!}{
    \begin{tabular}{cc}
    \hline\hline
    \multicolumn{2}{c}{\hii Regions} \\
    \hline
       Model reference &  \cite{Perez-Montero2014}\\
       log(U) & (-4.00, -1.50) in steps of $0.25$\\
       12+log(O/H) & (7.1, 9.1) in steps of $0.1$ \\
       Constraints & C tied to O \\
       log(N/O) & (-0.825, -0.625) in steps of $0.125$ \\
    \hline
    \multicolumn{2}{c}{Planetary Nebulae} \\
    \hline
         Model reference & \cite{Delgado-Inglada2014}\\
         Central star L/$(10^2 \, L_{\odot})$ &  $(2, 10, 30, 56, 100, 178)$ \\
         SED &  T.Rauch (TR) models \citep{Rauch2003}\\
         Central star $T_{\rm eff}/(10^3$ K) & ($50 -180$) in steps of $25$\\
         Dust depletion &  Yes\\
         Density [$\rm cm^{-3}$] &  $(30, 100, 300)$\\
         Radius [$\rm 10^{17}$ cm] &  $(1,3)$\\
         12 + log(O/H)&  8.34\\
         Model set &  matter bounded models (M40)\\
    \hline
    \multicolumn{2}{c}{Shock Regions} \\
    \hline
         Model reference &  \cite{Allen2008}\\
         Abundances & LMC  \citep{Russell1992}\\
         Magnetic field [$\mu$G] & $(0.0001, 0.5, 1, 2, 3.23, 4, 5, 10 )$ \\
         Shock velocities [$\rm km \, s^{-1}$] & (10 $-$ 100) in steps of $25$\\
         Pre-shock density [$\rm cm^{-3}$] & $1$\\
         Model set & Shock w/o precursor \\
    \hline
    \multicolumn{2}{c}{DIG} \\
    \hline
       Model reference &  \cite{Flores-Fajardo2011}\\
       $\log(U)$ & (-4.0 $-$ -3.0) in steps of $0.1$\\
       $\phi_{OB}$  &  (3.50 $-$ 5.00) in steps of $0.25$\\
       log(N/O) & -1.1, -0.9, -0.8, -0.7, -0.6\\
       12 + log(O/H) & (7.7 $-$ 9.3) in steps of $0.1$\\
    \hline
    \end{tabular}
    }
    \label{tab:models}
\end{table}

\section{Classification of integrated nebular spectra}
\label{sec:AppB}

\subsection{Integrated emission-line region spectra} \label{sec:regions}

Following Feltre et al. (in prep.), we identified line-emitting regions of interest (\hii\ regions, PNe, and SNR) in the M33 MUSE datacube according to a two-step process. First, the hierarchical tree-based Python package \texttt{astrodendro} was used on the combined \ha+\oiii$\lambda5007$ line emission map. 
The \texttt{astrodendro} hyperparameters were chosen to produce a segmentation mask matching visual expectations and detections of individual nebulae from existing catalogues. 
We used slightly less restrictive parameters than those adopted in Feltre et al. (in prep.), which focused primarily on \hii\ regions, in order to maximise the detection of SNR and  PNe. 

In a second step, the final segmentation mask was obtained by deblending adjacent regions using the watershed algorithm and requiring a minimum number of 100 pixels per region. This process yielded a total of 179 nebular regions. Integrated spectra were extracted from the MUSE data cube by summing the flux within the mask of each region. 
For these, we adopt the classification of these regions into \hii\ regions, PNe, and SNRs as presented in Feltre et al. (in prep.), which is based on traditional diagnostics and takes into account literature studies of PNe and SNRs.

In addition to these 179 regions, we considered five PNe from the literature \citep{Ciardullo2004} that were not identified by \texttt{astrodendro} and extracted spectra from apertures of $2"$ around them. 
Furthermore we extracted spectra (which we label as DIG) from 16 apertures of $2"$ in the field, specifically chosen to avoid correspondence with bright \ha, \oiii, or \sii\ emission, as well as known ionising stars \citep{Williams2021}.
In total, we obtained 200 integrated spectra.
The average S/N measured on the \ha\ line for all the 200 spectra is $\sim370$.

\subsection{The `integrated nebulae' ML model}
\label{sec:tailoring_integrated}

We build a separate ML model to classify the spectra of the integrated regions in M33. Since these are very high S/N spectra (average of $\gtrsim300$ on \ha), there is a significant stellar continuum contribution. 
We find that in this case the continuum affects the models' performance. In particular, the fraction of misclassified instances decreased by $\sim30\%$ when classifying spectra after continuum subtraction.
We therefore subtract the continuum from the data using \texttt{pPXF} \citep{Cappellari2003} and a set of eMILES simple stellar population models \citep{Vazdekis2016}.

The generation of the mock spectra follows the same steps as those described in Sec. \ref{sec:mlspec}. 
Since the real integrated spectra have very high S/N, we train the network with S/N=370, the average value for the integrated spectra.
We also evaluate the extinction affecting the integrated spectra by measuring the Balmer decrement from the \ha/\hb\ flux ratio for all the spectra and replicate its distribution in the synthetic dataset, similarly to the procedure described in Sec. \ref{sec:tailoring_spaxels}.

We test the model and find an average median error on the prediction of each class' contribution of $\sim8\%$. 
The measured f1-score is equal to 0.89, considering only the test instances with the prevalent class contributing to at least 50$\%$ of the spectrum and comparing that class with the dominant one predicted by the network. This performance is compatible with that of the previously described multi-label cases (Sec. \ref{sec:mlspec}), considering that the S/N in this mock dataset is five times higher. 

\subsection{Classification of integrated spectra}\label{sec:integrated}

\begin{figure}
    \centering
    \includegraphics[width=0.9\linewidth]{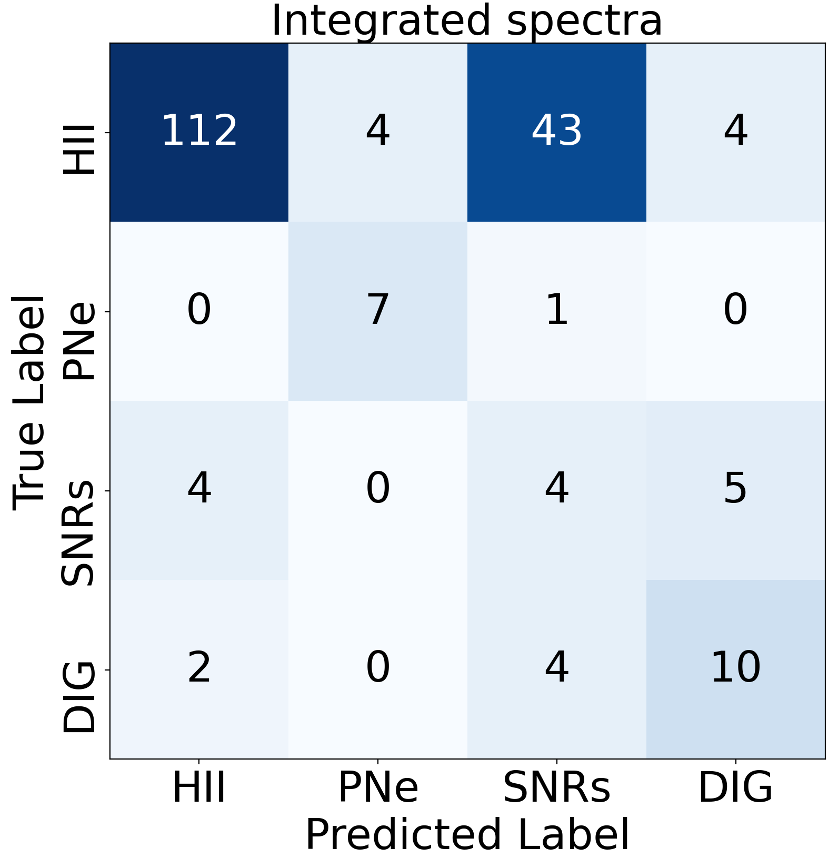}
    \caption{Confusion matrix comparing the traditional classification by Feltre et al. (in prep.) and the most probable class predicted by the neural network for the integrated spectra of the nebulae of M33.}
    \label{fig:cmmldend}
\end{figure}

\begin{figure}
    \centering
    \includegraphics[width=1\linewidth]{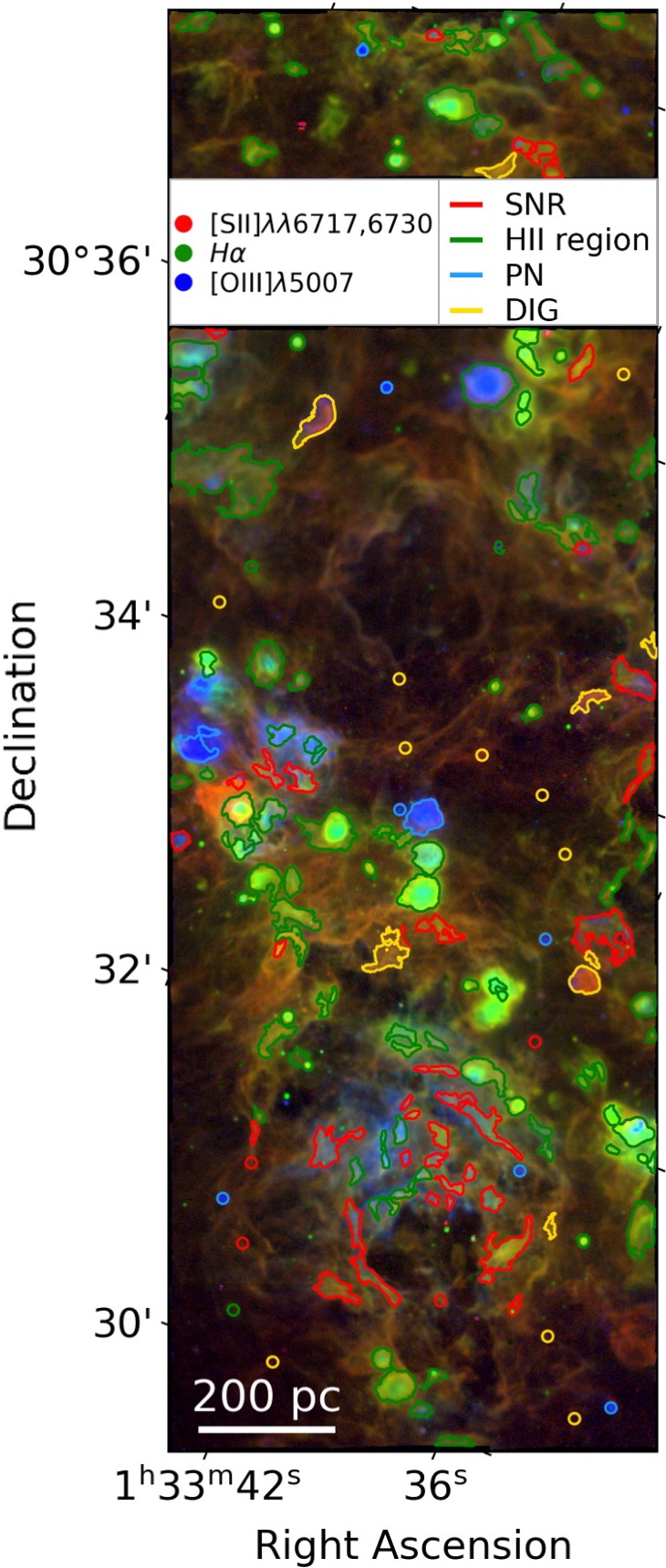}
    \caption{Map of the observed field of M33 showing the regions for the extraction of the integrated spectra and their neural network classification. We show the RGB map of three key emission lines for the entire observed field of M33, with the red scale representing the \siit flux, the blue the \oiii$\lambda5007$ flux, and the green the \ha\ flux.
    On top of this, we mark the contours of the regions of spectrum extraction coloured based on the model's classification: green for \hii region, blue for PNe, red for SNRs, and gold for DIG.
    }
    \label{fig:dendclassimg}
\end{figure}

In this section, we test the classification of artificial neural networks on the integrated spectra of the M33 emission-line regions extracted from the dendrograms, using the tailored model presented in Sec. \ref{sec:tailoring_integrated}.
As \quotes{ground truth} labels for emission-line regions in the M33 cube, we consider the classification carried out by Feltre et al. (in prep.) using traditional methods. In particular, they relied on the extensive literature classification of nebulae and ionising sources in M33, including catalogues of PNe \citep{Ciardullo2004}, SNRs \citep{Lee2014, Long2018}, Wolf-Rayet stars \citep{Neugent2011} and their associated nebulae \citep{Tuquet2024}, and the position of hot young stars selected from the HST UV-optical-IR photometry presented in \cite{Williams2021}.
On the other hand, the DIG labelled spectra were specifically extracted (Appendix \ref{sec:regions}) to avoid any of the other types of nebula from regions without \cite{Williams2021} ionising stars and bright \ha, \sii, or \oiii\ emission.

We now compare this classification with that of the neural network for all the 200 spectra.  
To evaluate the network's performance we create confusion matrices comparing the traditional classification of each spectrum with the most probable class the model predicts for each spectrum.
The number of spectra for each class is very inhomogeneous, with the majority of regions classified as \hii regions. 
From the confusion matrix (Fig.\ref{fig:cmmldend}), we observe that the fraction of correctly classified dendrograms is 68\% for those identified as \hii regions, 87\% for those identified as PNe, and 62\% for those identified as DIG. However, of the SNR spectra, only 30\% were classified as such.
Considering the integrated spectra whose classification does not correspond to the one obtained with traditional diagnostics, we notice the recurring confusion between SNR, DIG, and \hii region spectra, similar to that present also in the mock spectra classification (see Fig. \ref{fig:cm_all}).

In Fig. \ref{fig:dendclassimg} we present a map of the regions from which the integrated spectra have been extracted, colour coded based on the classification assigned by the network (green for \hii regions, blue for PNe, green for SNRs and gold for DIG).
The majority of dendrograms identified as \hii regions by traditional diagnostics but classified as SNRs or DIG by the neural network, exhibit comparatively higher fluxes of \oiii$\lambda5007$ or \siit\ compared to the other \hii regions, identified by their hosting of ionising stars.
Additionally, the network better identifies those \hii regions with a prevalent \ha\ emission (compared to other lines), thus appearing red in the emission-line map in Fig. \ref{fig:dendclassimg}. 
The misclassified PN occurs near a different type of bright nebula (specifically a SNR), which indeed leads the network towards an incorrect SNR classification. This PN is also misclassified by our single spaxel model (see Fig. \ref{fig:pne_cutouts}, panel h), which predicts a higher \hii region probability.

Of all the \hii regions, four are classified as PNe. Similarly to the case of the single spaxel model (Sect. \ref{sec:ifsclassif}), they are all characterised by bright \oiii\ emission.
Furthermore, the many \hii regions classified as SNR or DIG have low \ha\ emission, lie in lower S/N areas, or do not directly host an ionising star, therefore potentially challenging the validity of the label from Feltre et al. (in prep.).
Conversely, the four SNRs identified as \hii regions are characterised by brighter \ha\ emission with respect to the other dendrograms corresponding to SNRs found in the literature. 
Additionally, as in the single spaxel model (Sect. \ref{sec:ifsclassif}), there is a significant \quotes{confusion} between SNRs and DIG spectra (see Fig. \ref{fig:cmmldend}).
However, as discussed in Sec. \ref{sec:ifsclassif}, the model is not able to rely on the \oi$\lambda6300$ line, which, according to both \cite{Kopsacheili2020} and our neural network models (Sec. \ref{sec:actmax}), is crucial in identifying and distinguishing SNRs.

Finally, for many of the correctly identified integrated spectra, the second most probable classification is DIG.
However, for the majority of the incorrectly classified instances, the second most probable class is the correct one.

\end{appendix}
 
\label{LastPage}
\end{document}